\documentclass[journal=jacsat,manuscript=article]{achemso}
\pdfoutput=1
\usepackage[version=3]{mhchem} 

\usepackage[xindy]{glossaries}
\newacronym{adsp}{ADSP}{ALCF Data Science Program}

\newacronym{ar}{AR}{Average Recall}

\newacronym{ap}{AP}{Average Precision}

\newacronym{tdw}{TDW}{ThreeDWorld}

\newacronym{dwt}{DWT}{Discrete Wavelet Transform}

\newacronym{cv}{CV}{Computer Vision}

\newacronym{detr}{DETR}{DEtection TRansformer}

\newacronym{gpu}{GPU}{Graphical Processing Unit}

\newacronym{rnn}{RNN}{Recurrent Neural Network}

\newacronym{hmm}{HMM}{Hidden Markov Model}

\newacronym{ba}{BA}{Brodmann Area}

\newacronym{drl}{DRL}{Deep Reinforcement Learning}

\newacronym{sft}{SFT}{Semantic Folding Theory}

\newacronym{ca}{CA}{cornu Ammonis}

\newacronym{ai}{AI}{Artificial Intelligence}

\newacronym{bptt}{BPTT}{Backpropagation Through Time}

\newacronym{wer}{WER}{Word Error Rate}

\newacronym{tca}{TCA}{Temporal Credit Assignment}

\newacronym{lsa}{LSA}{Latent Semantic Analysis}

\newacronym{bert}{BERT}{Bidirectional Encoder Representations from Transformers}

\newacronym{gpt}{GPT}{Generative Pre-trained Transformer}

\newacronym{ds}{DS}{Distributional Semantic}

\newacronym{agl}{AGL}{Artificial Grammar Learning}

\newacronym{cstc}{CSTC}{Cortico-Striato-Thalamo-Cortical}

\newacronym{ann}{ANN}{Artificial Neural Network}

\newacronym{hpsg}{HPSG}{Head-Driven Phrase Structure Grammar}

\newacronym{dl}{DL}{Deep Learning}

\newacronym{rf}{RF}{Random Forest}

\newacronym{ml}{ML}{Machine Learning}

\newacronym{erps}{ERPs}{Event Related Potentials}

\newacronym{uf}{UF}{Unification Framework}

\newacronym{nlp}{NLP}{Natural Language Processing}

\newacronym{nlu}{NLU}{Natural Language Understanding}

\newacronym{mpf}{MPF}{Memory-Prediction Framework}

\newacronym{lifg}{LIFG}{Left Inferior Frontal Gyrus}

\newacronym{cpu}{CPU}{Central Processing Unit}

\newacronym{alcf}{ALCF}{Argonne Leadership Computing Facility}

\newacronym{anl}{ANL}{Argonne National Laboratory}

\newacronym{dmn}{DMN}{Deep Maxout Network}

\newacronym{dnn}{DNN}{Deep Neural Network}

\newacronym{dna}{DNA}{Deoxyribonucleic acid}

\newacronym{snr}{SNR}{Signal-to-Noise Ratio}

\newacronym{cnp}{$C_{np}$}{concentration of nanospheres}

\newacronym{dnp}{$D_{np}$}{diameter of nanopore}

\newacronym{rpd}{RPD}{Relative Percent Difference}

\newacronym{sgd}{SGD}{Stochastic Gradient Descent}

\newacronym{gsom}{GSOM}{Growing Self Organizing Map}

\newacronym{se}{SE}{Standard Error}

\newacronym{bn}{B-Net}{Bi-path Network}

\newacronym{nt}{NT}{Nanopore Translocation}

\newacronym{cdbn}{CDBN}{Convolutional Deep Belief Networks}

\newacronym{cnn}{CNN}{Convolutional Neural Network}

\newacronym{rms}{RMS}{Root-Mean-Square}

\newacronym{mse}{MSE}{Mean-Square Error}

\newacronym{mlp}{MLP}{MultiLayer perceptron}

\newacronym{fair}{FAIR}{Facebook's AI Research}

\newacronym{resnet}{ResNet}{Residual Neural Network}

\newacronym{rl}{RL}{Reinforcement Learning}

\newacronym{rlr}{RLR}{regularLayerResponse}

\newacronym{ipc}{IPC}{Inter-Process Communication}

\newacronym{pca}{PCA}{Principal Components Analysis}

\newacronym{adaboost}{AdaBoost}{Adaptive Boosting}

\newacronym{stl}{STL}{Standard Template Library}

\newacronym{ssom}{SSOM}{Static Self-Organizing Map}

\newacronym{dsom}{DSOM}{Dynamic Self-Organizing Map}

\newacronym{csom}{CSOM}{Complex Self-Organizing Map}

\newacronym{el}{EL}{Encoder Layer}

\newacronym{elslc}{EL SLC}{Encoder Layer with Stripped Lateral Connections}

\newacronym{lstm}{LSTM}{Long Short-Term Memory}

\newacronym{knn}{k-NN}{k-nearest neighbor}

\newacronym{lsm}{LSM}{Least Squares Method}

\newacronym{dt}{DT}{Decision Tree}

\newacronym{smpi}{SMPI}{single-molecule protein identification}

\newacronym{bdm}{BDM}{Bayesian Decision Making}

\newacronym{dtw}{DTW}{Dynamic Time Warping}

\newacronym{rt}{RT}{Recognition Tunnelling}

\newacronym{cusum}{CUSUM}{Cumulative Sums}

\newacronym{mos2}{MoS$_2$}{molybdenum disulfide}

\newacronym{omp}{OpenMP}{Open Multi-Processing}

\newacronym{mrstsa}{MRSTSA}{Multiresolution Spectro-Temporal Sound Analysis}

\newacronym{mfb}{MFB}{Mel Filter Bank}

\newacronym{strf}{STRF}{Spectro-Temporal Receptive Field}

\newacronym{fft}{FFT}{Fast Fourier Transform}

\newacronym{ffnn}{FFNN}{Feed-forward Fully connected Neural Network}

\newacronym{htm}{HTM}{Hierarchical Temporal Memory}

\newacronym{cla}{CLA}{Cortical Learning Algorithm}

\newacronym{a1}{A1}{Primary Auditory Cortex}

\newacronym{stg}{STG}{Superior Temporal Gyrus}

\newacronym{cc}{CC}{Cortical Column}

\newacronym{cstm}{CSTM}{Cortical Spectro-Temporal Model}

\newacronym{pp}{PP}{Predictive Population}

\newacronym{pg}{PG}{Populations' Group}

\newacronym{tip}{TIP}{Temporal Integration Population}

\newacronym{sp}{SP}{Simple Population}

\newacronym{kfm}{KFM}{Kohonen Feature Map}

\newacronym{asts}{aSTS}{anterior Superior Temporal Sulcus}

\newacronym{psts}{pSTS}{posterior Superior Temporal Sulcus}

\newacronym{cl}{CL}{Cortical Layer}

\newacronym{mgb}{MGB}{Medial Geniculate Body}

\newacronym{ic}{IC}{Inferior Culliculus}

\newacronym{ps}{PS}{Population Structure}

\newacronym{ltp}{LTP}{Long-Term Potentiation}

\newacronym{ltd}{LTD}{Long-Term Depression}

\newacronym{igr}{iGluRs}{ionotropic Glutamate Receptors}

\newacronym{mgr}{mGluRs}{metabotropic Glutamate Receptors}

\newacronym{ids}{IDS}{infant-directed speech }

\newacronym{ads}{ADS}{adult-directed speech }

\newacronym{em}{EM}{Expectation Maximization}

\newacronym{mog}{MOG}{Mixture of Gaussians}

\newacronym{mfcc}{MFCC}{Mel Frequency Cepstral Coefficients}

\newacronym{fcm}{FCM}{Fuzzy c-Means}

\newacronym{som}{SOM}{Self Organizing Map}

\newacronym{vot}{VOT}{Voice Onset Time}

\newacronym{vl}{VL}{Vowel Length}

\newacronym{mmn}{MMN}{Mismatch Negativity}

\newacronym{eeg}{EEG}{Electroencephalograph}

\newacronym{meg}{MEG}{Magnetoencephalograph}

\newacronym{oop}{OOP}{Object-oriented programing}

\newacronym{dft}{DFT}{Discrete Fourier Transform}

\newacronym{mfe}{MFE}{Massive Firing Event}

\newacronym{std}{STD}{Standard Deviation}

\newacronym{stdp}{STDP}{Spike-timing dependent plasticity}

\newacronym{sopm}{SOPM}{Self Organizing Predictive Map}

\newacronym{hl}{HL}{Hierarchical Layers}

\newacronym{si}{SI}{Supporting Information}

\newacronym{cvcv}{CVCV}{consonant-vowel-consonant-vowel}

\newacronym{vcvc}{VCVC}{vowel-consonant-vowel-consonant}

\newacronym{nn}{NN}{Neural Network}

\newacronym{sdr}{SDR}{Sparse Distributed Representation}

\newacronym{nmda}{NMDA}{N-Methyl-D-aspartic acid}

\newacronym{lts}{LTS}{learning time-series shapelets}

\newacronym{svm}{SVM}{Support Vector Machine}

\newacronym{libsvm}{LIBSVM}{Library for Support Vector Machine}

\newacronym{liblinear}{LIBLINEAR}{Multi-core Support Vector Machine}

\newacronym{timit}{TIMIT}{Acoustic-Phonetic Continuous Speech Corpus}

\newacronym{puc}{PUC}{Positive Unlabeled Classification}

\newacronym{hpc}{HPC}{High Performance Computing}

\newacronym{mpi}{MPI}{Message Passing Interface}

\newacronym{tcs}{TCS}{Theory and Computing Sciences}

\newacronym{atpesc}{ATPESC}{Argonne Training Program on Extreme-Scale Computing}

\newacronym{api}{API}{Application Program Interface}

\newacronym{tpp}{TPP}{Truncated Pyramid Nanopore}

\newacronym{psd}{PSD}{Power Spectrum Density}

\newacronym{wht}{WHT}{Walsh-Hadamard Transform}

\newacronym{dbc}{DBC}{second-order-differential-based calibration}

\newglossaryentry{deltaOmega}
{
  name = {\ensuremath{\Delta \omega_{ij}}},
  description = {A real value which represents the modification that must be made to the synaptic weight that belongs to unit $i$ in its input $j$},
  sort  = deltaOmega
}

\newglossaryentry{eta}
{
  name = {\ensuremath{\eta}},
  description = {Learning Rate. A multiplicative factor that changes the rate to which the synaptic weights must be modified},
  sort  = eta
}

\newglossaryentry{lambda}
{
  name = {\ensuremath{\Lambda(i,i^*)}},
  description = {The Neighborhood Function which modulates the synaptic weight modification as a function of the distance between the winner unit $i^*$ and another unit $i$ in the output space},
  sort  = lambda
}

\newglossaryentry{inputVector}
{
  name = {\ensuremath{\boldsymbol{\xi}}},
  description = {Input Vector of \ac{som}},
  sort  = inputVector
}

\newglossaryentry{synapticVector}
{
  name = {\ensuremath{\boldsymbol{\omega}_i}},
  description = {Synaptic Vector of output unit $i$},
  sort  = synapticVector
}

\newglossaryentry{inputComponent}
{
  name = {\ensuremath{\xi_j}},
  description = {Component $j$ of the input vector \ac{inputVector}},
  sort  = inputComponent
}

\newglossaryentry{synapticWeight}
{
  name = {\ensuremath{\omega_{ij}}},
  description = {Synaptic Weight $j$ in the output unit $i$},
  sort  = synapticWeight
}

\newglossaryentry{vectorPosition}
{
  name = {\ensuremath{\boldsymbol{r}_i}},
  description = {Vector Position of the unit $i$ into the output space},
  sort  = vectorPosition
}

\newglossaryentry{widthParameter}
{
  name = {\ensuremath{\sigma}},
  description = {Width Parameter that modifies the lateral influence in the learning process into the output space},
  sort  = widthParameter
}

\newglossaryentry{outputUnit}
{
  name = {\ensuremath{O_i}},
  description = {Output Unit $i$ in \ac{som}},
  sort  = outputUnit
}

\newglossaryentry{neighborFiring}
{
  name = {\ensuremath{\lambda}},
  description = {Neighbor Firing. This is the maximum distance from the input vector $\boldsymbol{\xi}$ admitted in an output unit $O_i$ to fire},
  sort  = neighborFiring
}

\newglossaryentry{predictiveDeltaOmega}
{
  name = {\ensuremath{\Delta \tilde{\omega}_{i^{**}i}}},
  description = {A real value which represents the modification that must be made to the predictive synaptic weight that belongs to unit $i$ in the connection with the last winner unit $i^{**}$},
  sort  = predictiveDeltaOmega
}

\newglossaryentry{predictiveEta}
{
  name = {\ensuremath{\tilde{\eta}}},
  description = {Predictive Learning Rate. A multiplicative factor that changes the rate to which the predictive synaptic weights must be modified},
  sort  = predictiveEta
}

\newglossaryentry{predictiveLambda}
{
  name = {\ensuremath{\tilde{\Lambda}(i,i^*)}},
  description = {The Predictive Neighborhood Function which modulates the predictive synaptic weight modification as a function of the distance between the winner unit $i^*$ and another unit $i$ in the output space},
  sort  = predictiveLambda
}

\newglossaryentry{predictiveWidthParameter}
{
  name = {\ensuremath{\tilde{\sigma}}},
  description = {Predictive Width Parameter that modifies the lateral influence in the predictive learning process into the output space},
  sort  = predictiveWidthParameter
}

\newglossaryentry{predictiveSynapticWeight}
{
  name = {\ensuremath{\tilde{\omega}_{i^{**}i}}},
  description = {Predictive Synaptic Weight of the output unit $i$ in connection with the last winner unit $i^{**}$},
  sort  = predictiveSynapticWeight
}

\newglossaryentry{firingPredisposition}
{
  name = {\ensuremath{FP_i}},
  description = {Firing Predisposition of unit $i$},
  sort  = firingPredisposition
}

\newglossaryentry{outputWinnerUnit}
{
  name = {\ensuremath{O_{i^*}}},
  description = {Output Winner Unit. This is the output unit which satisfies the rule $|\boldsymbol{\omega}_{i^*} - \ac{inputVector}| \leq |\ac{synapticVector} - \ac{inputVector}| \quad \quad \forall i$},
  sort  = outputWinnerUnit
}

\newglossaryentry{lastOutputWinnerUnit}
{
  name = {\ensuremath{O_{i^{**}}}},
  description = {Last Output Winner Unit. This is the output unit which satisfied the rule $|\boldsymbol{\omega}_{i^*} - \ac{inputVector}| \leq |\ac{synapticVector} - \ac{inputVector}| \quad \quad \forall i$ during the last activity executed in the \ac{sopm}},
  sort  = lastOutputWinnerUnit
}

\newglossaryentry{gnu_octave}
{
  name = {GNU Octave},
  description = {Scientific Programming Language},
  sort  = gnu_octave
}

\newglossaryentry{festival}
{
  name = {Festival Text to Speech},
  description = {The Festival Speech Synthesis System},
  sort  = {Festival}
}

\usepackage[pdftex,dvipsnames]{xcolor}

\author{Dario Dematties}
\affiliation{Instituto de Ciencias Humanas, Sociales y Ambientales, CONICET Mendoza Technological Scientific Center, Mendoza M5500, Argentina}
\altaffiliation{These two authors contributed equally}

\author{Chenyu Wen}
\altaffiliation{These two authors contributed equally}

\author{Mauricio David P\'erez}
\affiliation{Division of Solid-State Electronics, Department of Electrical Engineering, Uppsala University, SE-751 03 Uppsala, Sweden}

\author{Dian Zhou}
\affiliation{Department of Electrical and Computer Engineering, University of Texas at Dallas, Richardson, TX 75080, USA}

\author{Shi-Li Zhang}
\email{shili.zhang@angstrom.uu.se}
\phone{+46 701679347}
\affiliation{Division of Solid-State Electronics, Department of Electrical Engineering, Uppsala University, SE-751 03 Uppsala, Sweden}

\title[Nanopore sensing signals with bi-path network]
  {Deep learning of nanopore sensing signals using a bi-path network}


\begin{document}

\begin{tocentry}

\begin{figure}[H]
  \centering
  \includegraphics[width=8cm]{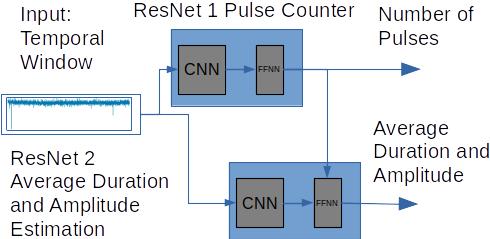}
   \label{fig:TOC}
\end{figure}

\end{tocentry}

\begin{abstract}
Temporary changes in electrical resistance of a nanopore sensor caused by translocating target analytes are recorded as a sequence of pulses on current traces. Prevalent algorithms for feature extraction in pulse-like signals lack objectivity because empirical amplitude thresholds are user-defined to single out the pulses from the noisy background. Here, we use deep learning for feature extraction based on a \gls{bn}. After training, the \gls{bn} acquires the prototypical pulses and the ability of both pulse recognition and feature extraction without \emph{a priori} assigned parameters. The \gls{bn} performance is evaluated on generated datasets and further applied to experimental data of DNA and protein translocation. The \gls{bn} results show remarkably small relative errors and stable trends. The \gls{bn} is further shown capable of processing data with a signal-to-noise ratio equal to one, an impossibility for threshold-based algorithms. The developed \gls{bn} is generic for pulse-like signals beyond pulsed nanopore currents. 
\end{abstract}

Nanopore sensing technology finds a wide scope of applications, including DNA sequencing \cite{jain2015improved}, protein profiling \cite{yusko2017real}, small chemical molecule detection \cite{borsley2018situ} and nanoparticle characterization \cite{lan2011nanoparticle}.
When analytes pass through a nanopore, characteristic pulses or spikes are generated on monitoring current traces \cite{Wen2020review}. Properties of the analytes, such as size, concentration, charge, dipole and shapes can be inferred from the amplitude, width (duration), frequency and waveform of such spikes \cite{yusko2017real, lan2011nanoparticle, houghtaling2019estimation, larkin2014high}. 
Traditional procedures in several different variants to recognise and extract translocation events, \emph{i.e}, spikes, from noisy current traces are typically based on a user-defined amplitude threshold as a criterion to separate the spikes from background noise fluctuations \cite{raillon2012fast, gu2015accurate}.
The flow of data processing for nanopore signals, as well as related algorithms, is a widely accepted establishment (see S.1 of \gls{si}).
The determination of spikes is, thus, highly dependent on how the threshold is defined. There is apparently an evident risk that this approach becomes subjective.
Although progress has been made in diminishing the subjectivity with empirical selection of the threshold, such as defining the threshold by referring to the background noise level, user intervention cannot be totally avoided \cite{zeng2019rectification}.
Therefore, conventional techniques for extracting features from raw data have been historically limited in its capacity.
To resolve this problem, an advanced algorithm based on a novel \gls{dl} architecture in the form of a \glsfirst{bn} is proposed in this work for spike recognition and feature extraction.
The \gls{bn} is capable of transforming directly raw data into an appropriate representation from which certain ending subsystems, such as a classifier, can detect patterns in the input. In other words, the three important features of the spikes, \emph{i.e.} amplitude, frequency and duration, can be extracted as a package solution covering the demands for an appropriate nanopore sensing technology.

\begin{figure}[H]
  \centering
  \includegraphics[width=11cm]{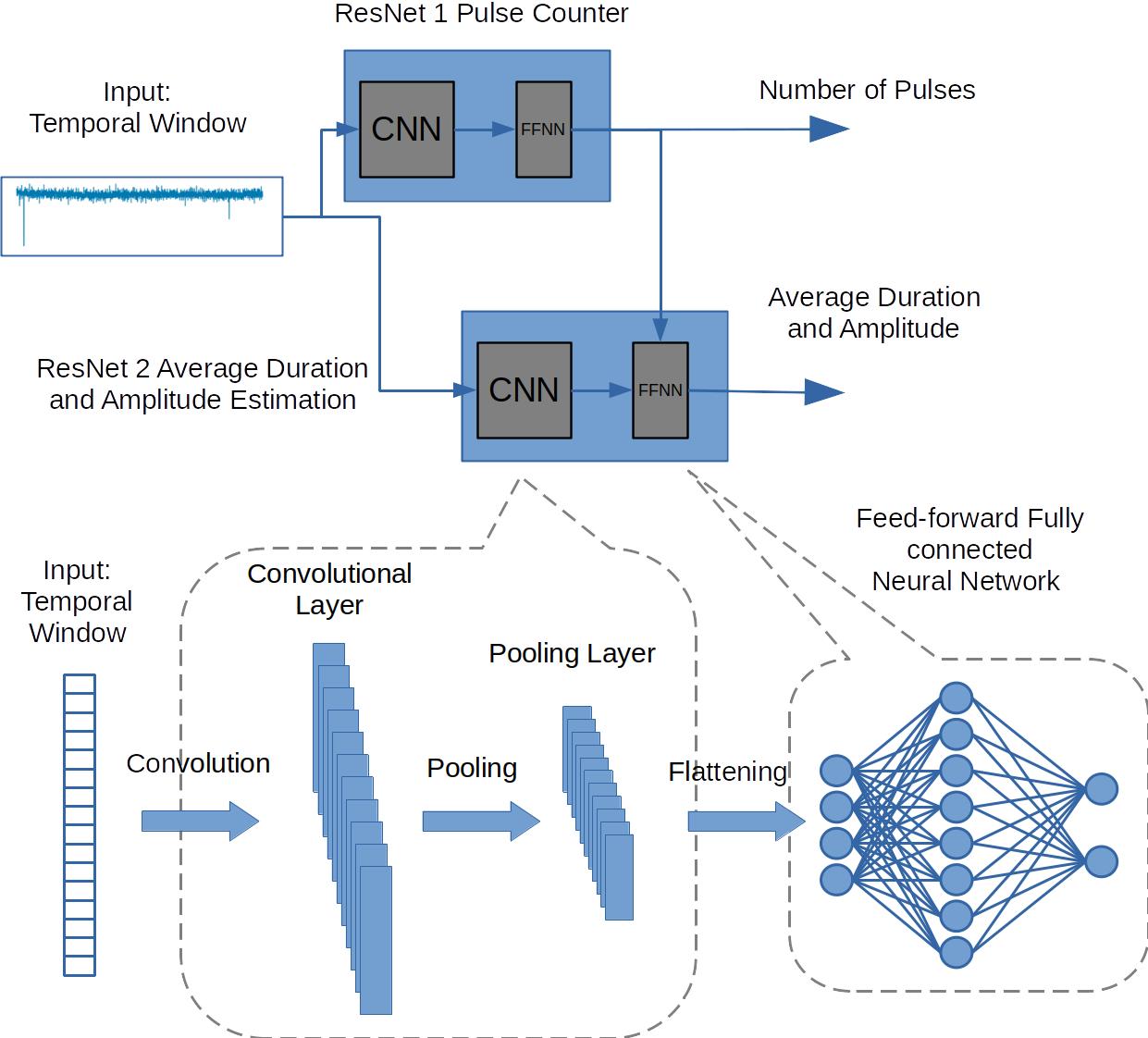}
  \caption{Architecture of the \gls{bn}. This novel network features a two-way architecture with two \glspl{resnet}.
  Each \gls{resnet} consists of a \gls{cnn} and a \gls{ffnn}.
  \gls{resnet} 1 predicts the number of pulses--or translocation events--in a temporal window. \gls{resnet} 2 forecasts the average translocation amplitude and duration of all the pulses within the same window. \gls{resnet} 1 also feeds its output, as an internal input, to the \gls{ffnn} of \gls{resnet} 2. The convolutional section in our implementation has been adapted to processing one-dimensional data. The fully connected architecture, on the other hand, outputs real valued predictions such as average amplitude and duration of the translocation events in a temporal window.}
  \label{fig:bipartite}
\end{figure}

The \gls{bn} is based on a highly consolidated \gls{dl} architecture, the \gls{resnet}, as depicted in Fig. \ref{fig:bipartite}.
As an integral part of the \gls{resnet} architecture, the \gls{cnn} can be utilised in any dimensional space although dimensions up to three are the most commonly used depending on the application.
For example, three-dimensional \glspl{cnn} are normally used for volumetric data in medical imaging \cite{10.1145/3065386}. On the other hand, two-dimensional \glspl{cnn} are the most popular for images and matrices \cite{10.1145/3065386}.
As is the case of the \gls{bn} introduced here, one-dimensional \glspl{cnn} have also been used for signals and time series such as in the case of automated detection of atrial fibrillation \cite{hsieh_detection_2020} and sleep arousal detection \cite{zabihi20191d}.

The \gls{bn} is first evaluated on artificially generated datasets. It is then employed for experimental data of $\lambda$-DNA and streptavidin translocation in solid-state nanopores.
Compared to its traditional algorithm counterparts, the \gls{bn} shows an outstanding performance from the perspective of robustness, objectivity and stability.
The developed \gls{bn} is in essence applicable for feature extraction of any pulse-like signals, \emph{e.g.}, transverse tunnelling current for single-molecule detection, spikes from neural cells and system, electro-cardio pulses, \emph{etc}.
It should be clarified that the \gls{bn} concept is designed for handling pulse-like signals. It is not meant for treating DNA/RNA sequencing data generated from a nanopore sequencer. Processing such sequencing data belong to a totally different field.

\section{Neural network architecture}

The holy grail in all \gls{dl} architectures is depth.
The rationale behind these successful methods is built on experimental evidence, which suggests that adding more layers to a \gls{dnn} provides us with more abstract output vectors that would better represent the hidden features from raw input data. Such rich features would finally allow us to better perform the task for which the network is trained. Yet, this desirable phenomenon has important limitations, since adding more layers to a network comes with penalties, such as vanishing and exploding gradients \cite{Hochreiter:91, Hochreiter:01book}.
\gls{resnet} is adopted in the \gls{bn}.
\gls{resnet} is an artificial \gls{dnn} that uses \glspl{cnn} and \emph{skip connections}, or \emph{shortcuts} that bypass some layers \cite{he2015deep, 7780459}.
The main motivation for bypassing/skipping layers is to avoid the problem of vanishing / exploding gradients.
The architecture of the \gls{bn} is visualised in Fig. \ref{fig:bipartite}.
It is a bi-path architecture composed of two parallel \glspl{resnet}.
Both \glspl{resnet} receive the same input that is simply a temporal window, segmented from a complete nanopore translocation current trace.
They also return predicted number of pulses and average amplitude and duration of such pulses in the window.

\begin{figure}[H]
  \centering
  \includegraphics[width=15cm]{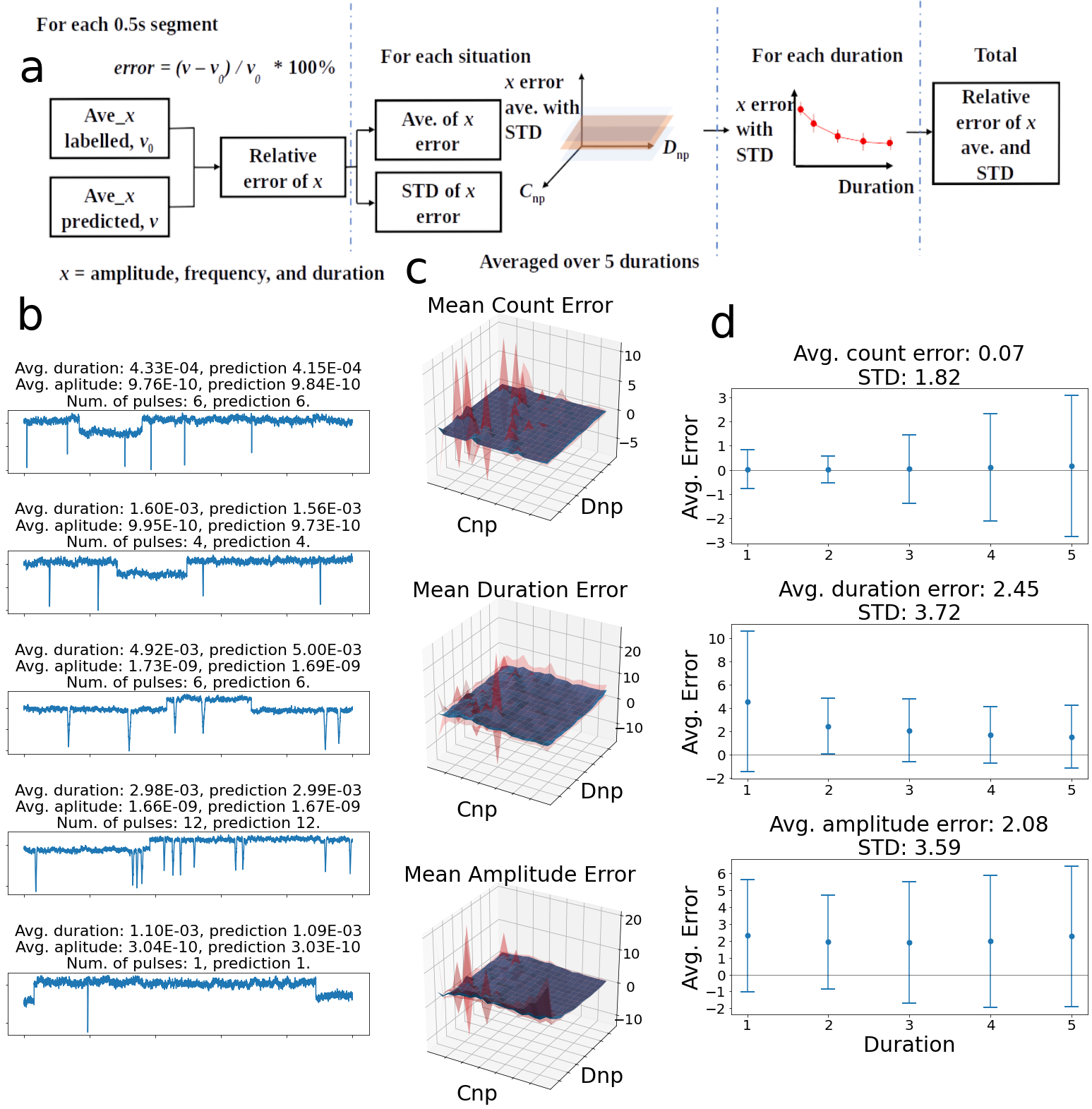}
  \caption{General output of the \gls{bn} and evaluation of the artificially generated test dataset with \gls{snr}=4.
  (a) Complete procedure of performance evaluation of the \gls{bn}. First, the relative errors between ground-truth values and predictions of the \gls{bn} are calculated. Second, the average error and \gls{std} are computed throughout the duration. Third, the average error and \gls{std} for each duration are calculated and finally the total average error and \gls{std} are obtained.
  (b) Example of five trace windows. The ground-truth feature values and the predictions of the \gls{bn} are listed above each window.
  (c) Surface plots showing the relative errors and their \glspl{std} for each situation vs \gls{cnp} and \gls{dnp}, averaged throughout the duration.
  (d) Average relative errors and \glspl{std} for each duration. Final values with total average errors and \glspl{std} are given above each figure.
  }
  \label{fig:validation}
\end{figure}

As shown in Fig. \ref{fig:bipartite}, \gls{resnet} 1 is assigned to predict the number of translocation spikes (pulses) in the window, while \gls{resnet} 2 forecasts the average amplitude and duration of all the pulses found in the same window.
\gls{resnet} 1 also provides its output as an internal input for the \glsfirst{ffnn} in \gls{resnet} 2.
Additional description of the \gls{bn} architecture is provided in \textbf{Methods} and S.2 of \gls{si}.

\section{Results and discussion}

\subsection{Training and validation}
\label{train-val}
\gls{resnet} 1 and 2 in Fig. \ref{fig:bipartite} were implemented by utilising another \gls{dnn}, \gls{resnet} 18
in the Pytorch \gls{dl} framework (https://www.pytorch.org) \cite{kuangliu_kuangliupytorch-cifar_2021}.
Since \gls{resnet} is originally designed for image processing in two dimensions, necessary modifications were implemented on the architecture to adapt it for the one-dimensional data.
Additionally, the last linear layer was replaced by a \gls{mlp} with two layers and the necessary adaptations were adopted to feed an extra internal input from the \gls{resnet} 1 output into the \gls{resnet} 2 \gls{ffnn}, in its final layer.
Finally, some batch normalisation mechanisms in certain layers of the network were replaced by the group normalisation strategy.
Our implementation is publicly available in {\textbf{\color{blue}[To be provided after the first submission]}}. 

The \gls{bn} was trained and validated using artificially generated data.
It is worth noting that the generated datasets are physics-based involving a set of well-established physical models for nanopore-based sensors. They include the nanopore resistance model \cite{Wen2017Rmodel}, spike generation model \cite{wen_current_2021}, and noise model \cite{Wen2017noise}, by entailing stochastic variations of corresponding parameters in accordance to the related physical mechanisms (\textbf{Methods} and S.3 of \gls{si}).
The \gls{bn} was subsequently evaluated on and applied to both artificial and experimental data (\textbf{Methods} and S.4 of \gls{si}).
Further, five different instances of the network were trained for five different \glspl{snr} in the artificial datasets. 
In all cases, smooth $l_1$-loss and \gls{sgd} optimisation were adopted.
Additional details about training, such as batch sizes, learning rate schedules, number of epochs, time consumed and curves of loss and errors for all \gls{bn} instances in this work can be found in S.5 of \gls{si}.

\subsection{Features extracted from generated datasets}

The general output of the \gls{bn} and the process of performance evaluation are depicted in Fig. \ref{fig:validation}.
The \gls{bn} receives a temporal window from a signal trace and returns a prediction of the average amplitude and duration of all the pulses as well as the number of pulses in the window (Fig.~\ref{fig:bipartite}). Typical windows with the ground-truth are shown in Fig. \ref{fig:validation}b along with the resultant predicted average values of amplitude and duration as well as the count of pulses for artificially generated traces with \gls{snr}=4.

\subsection{Performance evaluation}
\label{Perf_eva}

The procedure to performance-evaluate the \gls{bn} is shown in Fig. \ref{fig:validation}a
(more details in \textbf{Methods} and S.4 of \gls{si}).
The test dataset contains traces for different configurations (Fig. \ref{fig:validation}b) at specified values of \glsfirst{cnp} (correlated to translocation frequency), \glsfirst{dnp} (correlated to spike amplitude) and translocation dwell time (correlated to duration).
The relative error of each data window is calculated for each parameter (\emph{i.e.}, number of spikes, amplitude and duration) referred to the ground-truth recorded during the data generation, following the formula shown in Fig. \ref{fig:validation}a (details in \textbf{Methods}).
Then, the relative errors are averaged throughout the duration in each situation of \gls{cnp} and \gls{dnp} as displayed in Fig. \ref{fig:validation}c.
The blue surface corresponds to the average value while the translucent red spike-look surfaces over and beneath the error depict the \glsfirst{std} of the error.
The average errors and \glspl{std} for each duration in the dataset are plotted in Fig. \ref{fig:validation}d. 
Finally, total average errors and \glspl{std} are computed and shown above each plot
(complete evaluation results in S.4 of \gls{si}).

\begin{figure}[H]
  \centering
  \includegraphics[width=16cm]{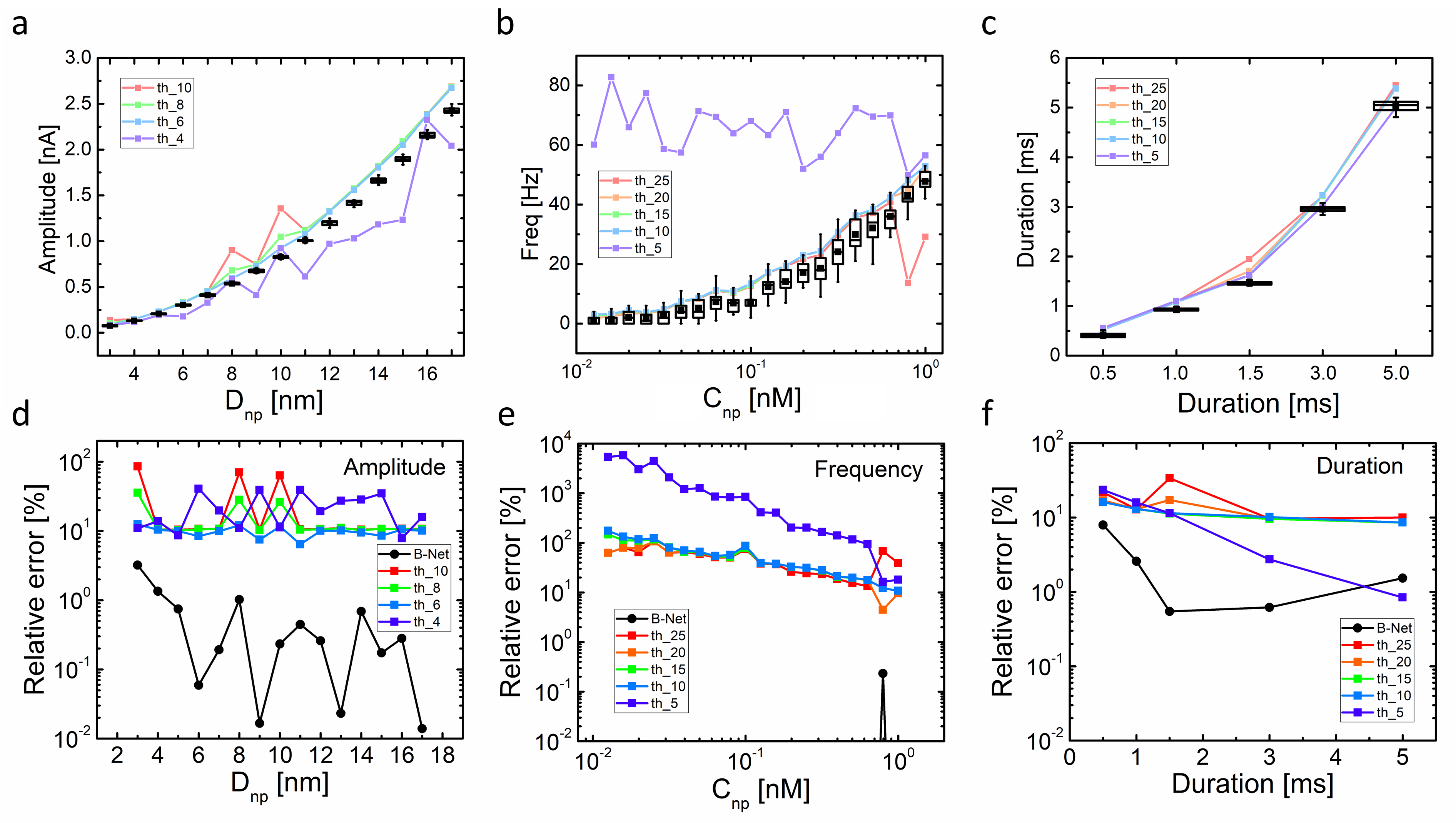}
  \caption{Comparison between the \gls{bn} and the traditional algorithm on the artificially generated test dataset with \gls{snr}=4.
  (a-c) Box charts for variation of spike amplitude, frequency and duration for translocating nanospheres with different \gls{dnp}, \gls{cnp} and duration, extracted using the \gls{bn} (black symbols).
  They are directly compared with average values for spike amplitude, frequency and duration extracted using the traditional algorithm with several different $n$ values in $th\_n$ (dot-on lines).
  Relative errors (whose expressions are given in Methods) for spike amplitude, frequency and duration predictions by the \gls{bn} and the traditional algorithm are compared in (d-f).
  }
  \label{fig:comparizon}
\end{figure}

The outputs and relative errors of the \gls{bn} are compared to those of the traditional algorithm in Fig. \ref{fig:comparizon}.
Different values of $n$ in $th\_ n$ denote the user-defined thresholds of the amplitude as a criterion to distinguish true translocation-generated spikes from fluctuations (noise) in a current trace.
Here, $th$ is the abbreviation of threshold and $n$ is defined as the number of multiples of the peak-to-peak value of the background noise (more details of the implementation of the traditional algorithm in \textbf{Methods}). The comparison has a focus on processing the artificially generated test dataset with \gls{snr}=4. 

The \gls{bn} and the traditional algorithm agree well in yielding similar variations of amplitude, frequency and duration (Fig. \ref{fig:comparizon}a-c).
Here, frequency is the ratio of the number of pulses predicted by the \gls{bn} and the time extension of a temporal window (0.5 s).
This computation was repeated for each window in the dataset where the statistical features were collected.
For the traditional algorithm, frequency is defined as the reciprocal of the time interval between the current and the inmediate next event \cite{zeng2019rectification}.

The average values of frequency, amplitude and duration features resulting from the traditional algorithm obviously depend on the selection of the amplitude threshold, as shown by the deviations among the dot-on lines in the figures.
Only if the threshold is properly selected, the traditional algorithm can return a reasonable result.
Small thresholds can lead to assignment of fluctuations as translocation spikes, thereby incorrectly increasing frequency count and decreasing both amplitude and duration numerical values.
On the other hand, large thresholds can give rise to missing translocation spikes, rendering misleadingly high average amplitude and duration.

Furthermore, the feature extraction resulting from the traditional algorithm is clearly less stable than that produced from the \gls{bn}.
For example, two small peaks appear in the $th\_10$ curve at \gls{dnp}=8 and 10 nm in Fig. \ref{fig:comparizon}a. Such instability arising from the oversensitivity of the traditional algorithm to data generation is obviously absent with the \gls{bn} approach.

The prediction errors for the different features for both algorithms, the \gls{bn} and the traditional, are compared in Fig. \ref{fig:comparizon}d-f.
Except for the prediction of duration in the duration=5 ms case (Fig. \ref{fig:comparizon}f), relative errors of the traditional method are notoriously larger than those of the \gls{bn}.
Even in the exception, the error of the \gls{bn} is still quite competitive.
Strikingly, the relative errors of the \gls{bn} for frequency count are four orders of magnitude smaller than those of the traditional algorithm for most \gls{cnp} values (Fig. \ref{fig:comparizon}e).
For amplitude (Fig. \ref{fig:comparizon}d) and duration (Fig. \ref{fig:comparizon}f), the relative errors of the \gls{bn} are up to three and two orders of magnitude, respectively, lower than those of the traditional algorithm.
In addition, the relative errors for the traditional algorithm are again highly dependent on the selection of the threshold, confirming the subjectivity of the algorithm.
Comprehensive comparisons of the amplitude, frequency and duration for each dataset, as well as their relative errors (S.6 of \gls{si} affirm the observations above.

\subsection{Signal-to-noise limitation}
\label{SNR_limit}

\begin{figure}[H]
  \centering
  \includegraphics[width=12cm]{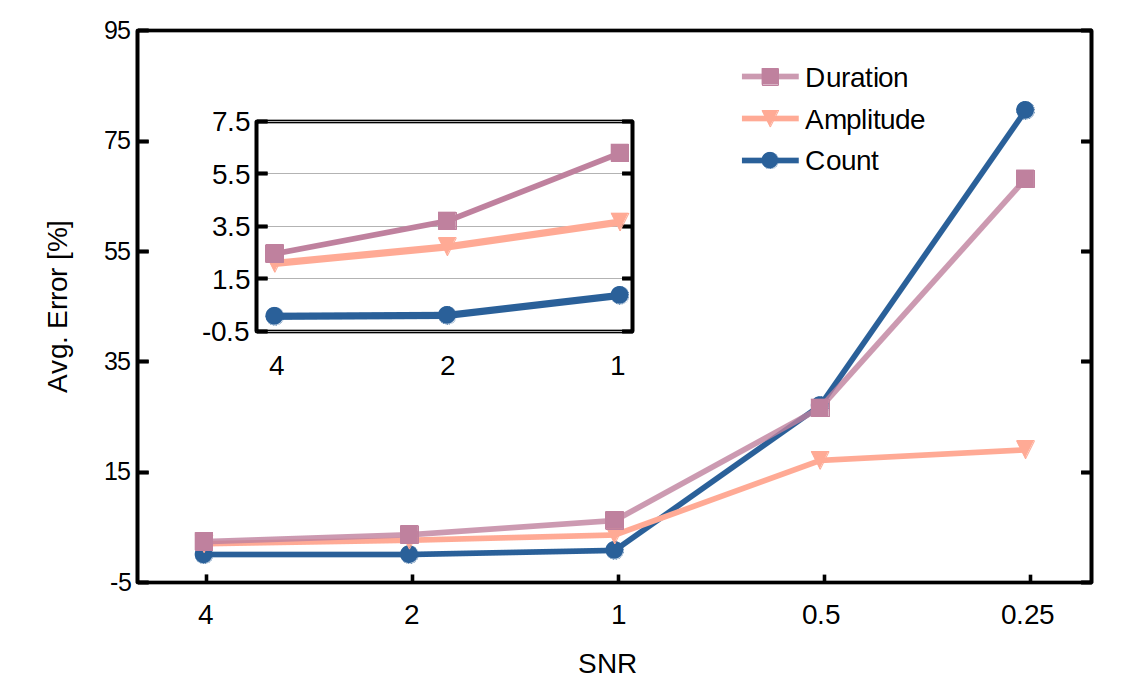}
  \caption{Total average relative errors of amplitude, frequency and duration \emph{vs.} \gls{snr} for the \gls{bn}.}
  \label{fig:error-noise}
\end{figure}

To explore the capability of the \gls{bn} in identifying signals from a noisy background of various noise levels, five different networks with the same architecture based on \gls{resnet} 18 were trained by datasets with \gls{snr} = 4, 2, 1, 0.5 and 0.25.
The performance of the \gls{bn} in response to the different noise levels is displayed in Fig. \ref{fig:error-noise} (complete evaluation in S.4 of SI).
Error values represent \gls{bn}'s relative errors in predicting the number of pulses (count) and the average amplitude and duration of all pulses in the temporal windows. As expected, errors increase with decreasing \gls{snr}.

Inferred from the steepest slope for the \emph{count} error in Fig. \ref{fig:error-noise}, \gls{resnet} 1, predicting number of pulses, is more affected than \gls{resnet} 2, predicting amplitude and duration, by the level of noise.
Similarly, the duration prediction is more susceptible than the amplitude counterpart.

The smallest error for count is produced by \gls{resnet} 1 with an average error of 0.067\% for \gls{snr}=4.
On the other hand, \gls{resnet} 2 produces an error of 2.5\% and 2.1\% in its prediction of duration and amplitude, respectively. 
For \gls{snr}=1, \gls{resnet} 1 is still the best performing part of the \gls{bn} with an average error of 0.86\% for count while \gls{resnet} 2 presents 6.3\% and 3.7\% of error for duration and amplitude, respectively.
Performance starts to degrade severely for \glspl{snr} below 1 with errors above 27\% (count) for \gls{resnet} 1 and above 26\% (duration) and 17\% (amplitude) for \gls{resnet} 2.

The performance of the \gls{bn} is compelling, given that it is almost impossible for the traditional algorithm to correctly recognise translocation spikes in a background noise that has a similar amplitude to the spikes, \emph{i.e.} \gls{snr}=1.
By using bigger architectures in combination with labelled generated datasets that are closer to real measured data it is possible to further reduce errors and facilitate correct processing of signals with even lower \gls{snr}.

In the \gls{bn}, the features of spikes are acquired by the algorithm during the training process. They contain as much of the original information as possible, indicating not only picking up obvious features considered in the traditional algorithm, such as amplitude and duration, but also noticing details in the pulses, such as the waveform.
These extra features assist the \gls{bn} to better appreciate the difference between a real translocation spike and a background noise peak, even when the two have the same amplitude with \gls{snr}=1. Furthermore, the decision process is flexible and probabilistic, indicating a powerful method with robust performance.
As the entire process minimizes the participation and intervention of users, it warrants maximum objectivity, see below.

\subsection{Objectivity analysis of experimental data}
\label{Exp_data}

\begin{figure}[H]
  \centering
  \includegraphics[width=16cm]{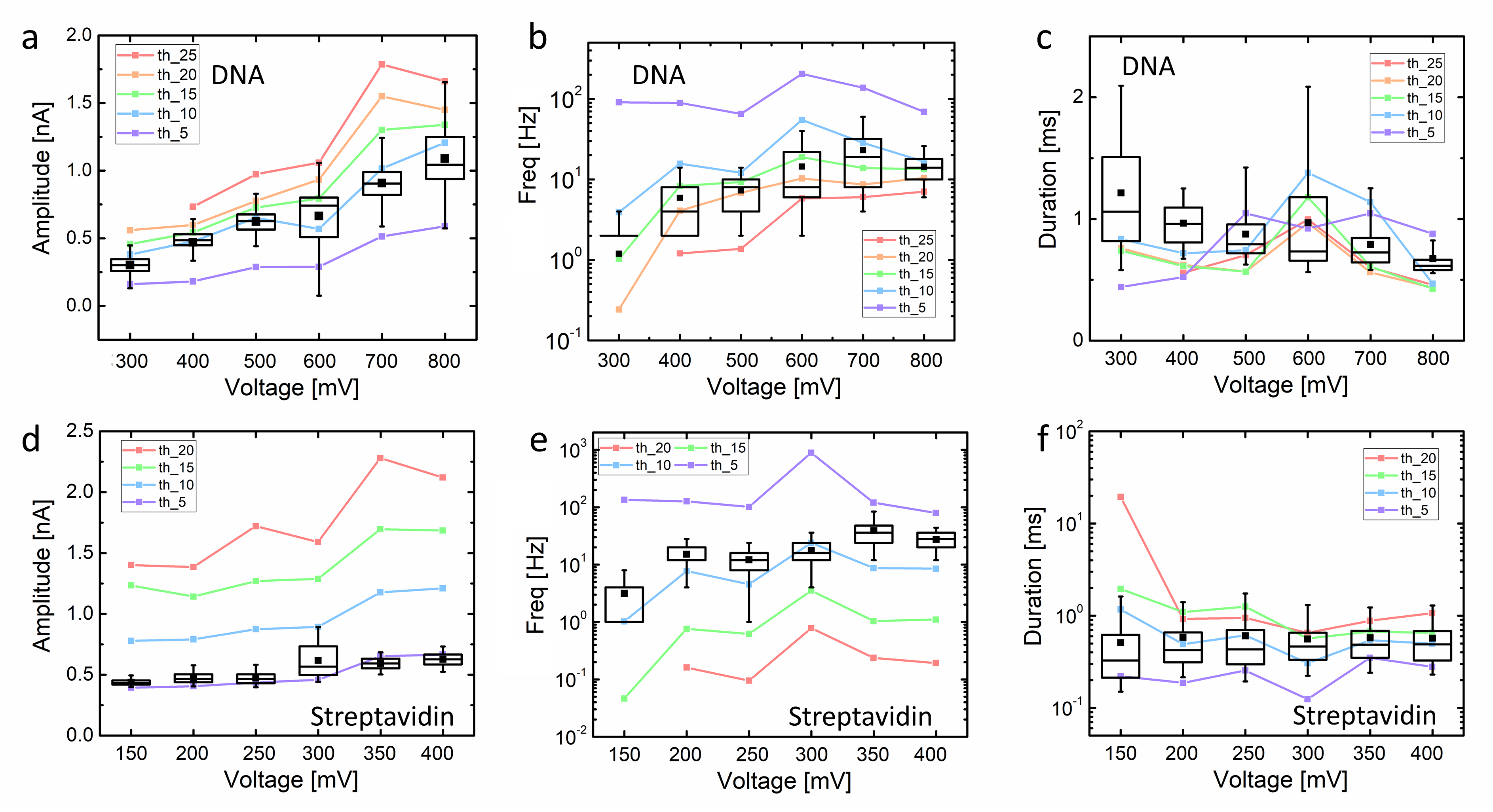}
  \caption{Signal processing of experimental data involving DNA and protein.
  Variation of spike amplitude, frequency and duration of $\lambda$-DNA (a-c) and \emph{streptavidin} (d-f) translocation with bias voltage.
  The box charts show the results from the \gls{bn}, while the dot-on lines represent the corresponding results from the traditional algorithm with different thresholds.}
  \label{fig:objective-comparizon}
\end{figure}

The \gls{bn} is also applied to two experimental datasets for nanopore translocation of $\lambda$-DNA and \emph{streptavidin} from our previous work \cite{zeng2019rectification}.
The outputs produced by the \gls{bn} and those produced by the traditional algorithm are compared in Fig. \ref{fig:objective-comparizon}a-c for the $\lambda$-DNA dataset.
There is a clear correlation regarding the variation in different features returned by both algorithms.
For the \gls{bn} outputs, both amplitude and frequency increase, while the duration decreases with bias voltage.
Data analysis shows a linear relationship for all (Fig. S20 of SI), which, to the first-order, is physically plausible.
In detail, the linear dependence of amplitude on bias voltage (Fig. \ref{fig:objective-comparizon}a and Fig. S20a) agrees with earlier reports \cite{fologea_slowing_2005,carlsen_interpreting_2014}. This is an expected relationship since for a blockage resistance generated by analyte translocation, a higher voltage induces proportionally a larger current. The translocation frequency also displays a linear dependence on voltage (Fig. \ref{fig:objective-comparizon}b and Fig. S20b), which indicates that the analyte capture probability is limited by diffusion, instead of by the energy barrier of the nanopore, since the latter would give rise to an exponential dependence on voltage \cite{wanunu_electrostatic_2010}. It has been shown experimentally that long double-stranded DNA (<104 bps), such as $\lambda$-DNA, usually display a diffusion limited capture behaviour \cite{grosberg_dna_2010}, in agreement with our finding here. Furthermore, a linear decrease of duration with increasing voltage (Fig. \ref{fig:objective-comparizon}c and Fig. S20c) is simply a result of increased electric field giving rise to a proportionally increased rate of translocation with an invariant ionic mobility \cite{li_distribution_2010}.
 
However, significantly larger dispersions and fluctuations in the output values are evident with the traditional algorithm whose outputs heavily depend on the subjective threshold determined by a user. 
The deviation among the results from the traditional algorithm with different thresholds is more significant than those for the generated data (Fig. \ref{fig:comparizon}a-c).
Hence, the features extracted from the traditional algorithm totally lose their objectivity. 

A similar comparison for streptavidin translocation (Fig. \ref{fig:objective-comparizon}d-e and Fig. S.7 of \gls{si}) displays also a
clear correlation of the variation of the outputs with bias voltage from both algorithms. Compared with the $\lambda$-DNA translocation data, the dependence of amplitude and duration on bias voltage is weaker though also linear, in agreement with other reports \cite{oukhaled_dynamics_2011,larkin_high-bandwidth_2014}.
The duration is insensitive to bias voltage, which could be related to the limited bandwidth at 10 kHz of the amplifier for data acquisition. When the translocation time is close to or shorter than the time constant defined by the cut-off frequency of the amplifier, the difference in the width of pulses is often smeared out \cite{pedone_data_2009,plesa_fast_2013}.
Furthermore, the dispersion presented by the traditional algorithm is worse than the one with the $\lambda$-DNA translocation data. This difference can be related to a lower \gls{snr} of the streptavidin translocation data. It is remarkable in Fig. \ref{fig:objective-comparizon}d-e that the amplitude and frequency predicted by the traditional algorithm are even more dependent on the choice of subjective voltage threshold.

\subsection{Conclusions}

The analyses and comparisons in preceding sections confirm eminently that the \gls{bn} meets the essential and critical requirements of being objective, avoiding subjective parameters determined by the user.
The adverse effects of a subjective and often blind input parameter adjustment are clearly appreciated in Figs. \ref{fig:comparizon} and \ref{fig:objective-comparizon}, where the predictions of the traditional algorithm sensitively depend on a threshold adjusted beforehand.
In contrast to the traditional algorithm replying on user-defined input parameters, the advantages of the \gls{bn} lie also in its clear, stable and consistent predictions  as well as negligibly small relative errors. All this is indicative of the robustness of the \gls{bn} in being able to analyse noisy data thereby easing the otherwise strict demands on the control of experimental conditions. The impressive performance of the \gls{bn} in singling out pulses from a noisy background with relative errors below 1\% for count and ~5\% for amplitude and duration for input data of \gls{snr}=1 provides a validating example.
Such performance is not anticipatable of traditional algorithms, since when thresholds are used for recognising translocation spikes, a noise peak, with similar amplitude, can easily be misclassified as a spike originated from a translocation event.
Furthermore, the bi-path architecture in the \gls{bn} assigns different categories of tasks (\emph{i.e.}, pulse count and average feature predictions) to distinct network branches, while the information processed in one branch (\emph{i.e.} the pulse counter) is used by the other to predict extra average features. This strategy is naturally in agreement with the architecture of the human brain.

Regardless all favourable features reviewed above, the \gls{bn} is a \gls{dl} based method.
As such, it is inherently a \emph{data hungry} strategy that works better when there are thousands, millions or even billions of training examples \cite{brown2020language}.
In problems with limited data sources, \gls{dl} is not an ideal solution. 
In the specific area of concern, real traces collected from nanopore translocation experiments could be abundant but they are not labelled.
Recruiting staff for labelling such data is not viable given the extension of the datasets needed to train the \gls{bn}.
Instead, we have generated our own artificial datasets in this work in order to train, validate and then put in use of the \gls{bn}.
The comparison experiments conducted against the traditional approaches by processing experimental traces clearly demonstrate that our generated datasets retain a high statistical correlation with the experimental traces collected in the laboratory.
Yet, it is imperative to clarify that beyond such favourable results, it is impossible to perfectly mimic the statistical distribution immersed in real traces.
General palliative methods to solve this problem are available to \gls{dl}.
There are pre-training stages of the networks with which looser requirements are demanded and smaller labelled datasets can later be used to fine-tune them.
Pre-training could be tackled using alternative generated labelled datasets.
In this work, we have shown the relevance of the datasets generated to train our \gls{bn}.
Our artificially generated datasets can also be used as a pre-training resource in the \gls{bn}.
Afterwards, adding labels to a much smaller dataset of traces collected in the laboratory can be a much more viable endeavour.
This could result in a highly qualified network, fine-tuned with real traces.
Such a network would show prominent performance differences from a network trained only using artificially generated datasets as the one introduced in this work.

In conclusion, our \gls{bn} algorithm is highly flexible to the input signal, and it is not limited to signals from nanopore sensors.
A myriad of pulse-like signals, found in biotechnology, medical technology, physical sciences and engineering, information and communication technology, environmental technology, \emph{etc.}, can be processed by implementing the robust and objective B-Net. Therefore, the B-Net is a generalisable and flexible platform owing to the flexibility of \gls{dl} strategies.

\section{Methods}

\subsection{Data preparation}
\label{dataset}

\subsubsection{Artificial data generation}
The artificially generated data is composed of three parts: 1) randomly appeared translocation spikes, 2) background noise and 3) baseline variations. The baseline current level, \emph{i.e.}, the open-pore current, is determined using the resistance model \cite{Wen2017Rmodel} , with given geometry properties, electrolyte concentration, and bias voltage (more details in S.3.1 of \gls{si}).
In this system, differently-sized nanospheres are used to represent analytes.
In the signal generation programme, the sampling rate is selected at 10 kHz to determine the time step of the signal.
According to our previous work \cite{Wen2018group}, the probability of appearance of translocation spikes at each time step is correlated to the concentration of nanospheres.
The amplitude of spikes is assigned by our translocation model based on the resistance change by steric blockage during the translocation \cite{Wen2016DNAsequencing}.
The waveform of translocation spikes is approximated using an asymmetrical triangle with adjustable ramping slopes (details in S.3.2 of \gls{si}).

According to the related studies, coloured Gaussian noise is adopted as the background noise \cite{smeets2008noise} whose power spectrum density is determined by our integrated noise model \cite{Wen2017noise}.
At frequencies below 5 kHz (confined by the 10 kHz sampling rate), the noise has four components: flicker noise, electrode noise, white thermal noise and dielectric noise, whose importance increases successively from low to high frequencies.
The related parameters are selected as the typical values of $SiN_x$ nanopores from our previous measurements \cite{Wen2017noise}.
The amplitude, reflecting the power, of the background noise can be tuned for datasets with different \gls{snr} (more details in S.3.3 of \gls{si}).

In addition, two kinds of variations of the baseline, \emph{i.e.}, sudden jumps and slow fluctuations, are introduced to represent the perturbation.
The former generates randomly appeared steps in the baseline to mimic the \emph{temporary adsorption-desorption} of some objects near and in the pore \cite{niedzwiecki2010single,Wen2020review} .
The latter simulates the instability of the nanopore, which can be caused by \cite{egatz2016future, freedman2014nonequilibrium} the fluctuation of pore membrane, dynamics of contamination, \emph{etc}. It is described by the superposition of eight terms of sine and cosine functions.
The probability of appearance of the perturbation steps, amplitude of the steps and the amplitude of the slow fluctuations are carefully tuned to approach the real common situations in experiments (more details in S.3.4 of \gls{si}).
Involving the baseline variations can largely enhance the robustness of our neural network.
Accustomed to complicated scenarios in the training process, the \gls{bn} can handle real measured data more competently.

In signal generation, both diameter and thickness of a nanopore are fixed to 20 nm. Typical experimental conditions are selected, including a bias voltage of 300 mV, a 100 mM KCl electrolyte and surface charge density of -0.02 C/m$^2$.
Three parameters in the signal generation programme, \emph{i.e.}, diameter of translocating nanospheres (aforementioned $D_{np}$), concentration of the nanospheres  (aforementioned $C_{np}$) and duration of translocation, are systematically varied for each dataset, since these three parameters have a direct correlation with the three most important features of the translocation signal, \emph{i.e.}, amplitude, frequency, and duration.
In each dataset, $D_{np}$ is varied from 3 nm to 17 nm, with a 1 nm step (\emph{i.e.}, 15 different values in total), while
$C_{np}$ is varied from 0.01 to 1 nM, changing in logarithmic scale (\emph{i.e.}, 20 different values in total).
The duration of translocation is directly assigned to 0.5, 1, 1.5, 3 and 5 ms (\emph{i.e.}, 5 different values in total).
To evaluate the performance of the \gls{bn} for different \gls{snr} conditions, \gls{snr} is varied from 0.25 to 4 (\emph{i.e.}, 5 different values in total).
It is worth noting that \gls{snr} is defined as the ratio of spike amplitude to the peak-to-peak value of the background noise. The peak-to-peak value of the Gaussian noise is estimated to be six times of its \gls{rms} value \cite{Wen2020review} , while the noise \gls{rms} can be calculated by the root square of the integration of the noise power spectrum density in the range of the bandwidth. All the data is generated using a homemade programme on MATLAB.

\subsubsection*{Experimental data}
In order to evaluate the performance of our algorithm on experimental data from laboratory, two groups of translocation experiments are implemented in \glspl{tpp}.
The \glspl{tpp} are formed in the single-crystal silicon layer of a silicon-on-insulator wafer. Their shape naturally adopts a truncated pyramid by exploiting the difference of wet etching rate of the $\langle 100 \rangle$ and $\langle 111 \rangle$ crystal orientations of single crystal silicon. Details about the nanopore fabrication can be found in our previous work \cite{zeng2019rectification} . Two different \glspl{tpp} were used in our experiments, one with a side length of 7.5 nm and another 16 nm, both in a 55 nm-thick silicon layer, for DNA and protein \emph{streptavidin} translocation respectively. 

$\lambda$-DNA and \emph{streptavidin} were selected as two typical examples of translocating analytes, representing two typical examples of long strand shape and sphere shape object, respectively.
$\lambda$-DNA has 48,502 nucleobase pairs in a double-stranded helix structure with a total length of about 16 $\mu$m \cite{tree2013dna} . \emph{Streptavidin} is sphere-like with a diameter of 6 nm \cite{yusko2017real} . They were purchased from Merck KGaA, Darmstadt, Germany, and used without further purification. The $\lambda$-DNA and \emph{streptavidin} were dispersed in 500 mM KCl electrolyte with a concentration of 78 pM and 84 nM, respectively.

The nanopore chip was mounted on a custom-made polymethyl methacrylate (PMMA) flow cell and sealed using two Polydimethylsiloxane (PDMS) O-rings (8 mm in inner diameter) on the two sides \cite{zeng2019rectification} . Two compartments both filled with KCl solution were separated by the chip and the only path of ionic current was through the nanopore.
A pair of Ag/AgCl pseudo reference electrodes (2 mm in diameter, Warner Instruments, LLC.) was used to apply a bias voltage and to collect the ionic current.
Electrical measurements were controlled using a patch clamp amplifier (Axopatch 200B, Molecular Device Inc.). During the translocation experiments, analyte dispersions were added to both compartments.
The ionic current was converted to the digital signal by Axon Digidata 1550A (Molecular Device LLC.) and recorded by software Axon pCLAMP 10 (Molecular Device LLC.).
The $\lambda$-DNA translocation was measured at 10 kHz sample rate with 2 kHz analogue bandwidth, while the \emph{streptavidin} translocation was detected at 20 kHz sampling frequency with a 10 kHz bandwidth.

\subsubsection{Traditional data processing method}
\label{traditional-method}
The reorganisation of translocation events by means of traditional data processing methods is based on the amplitude of spikes. Here, we use a homemade MATLAB programme to locate the translocation spikes in current traces and extract the three parameters: amplitude, translocation frequency, and duration.
In the programme, function \emph{findpeaks} is adopted with the \emph{MinPeakProminence} method.
An amplitude threshold is defined by the user regarding to the \gls{rms} of background noise level. In the following discussion, this threshold is tuned from 4 to 25 multiples of the background noise \gls{rms}, to demonstrate the dependence of the results on the threshold selection.

\subsubsection{Standard database}
A standard database for testing of the performance of translocation signal has been established on \textbf{{\color{blue}[the website of our database will be cited after the first submission]}}.
There are two categories of data, \emph{i.e.}, generated and experimental. For each dataset of the generated data, a current trace without background noise is also offered, apart from its counterpart trace with noise.
Furthermore, the spike information, including amplitude, start time, end time, and duration, is given as the standard answer for corresponding traces.
For the experimental data, the current traces are trimmed with the same length.
In addition, all the other important information, such as simulation and experiment conditions, is offered. This database is completely open to public and can be used to train, validate and test different algorithms, on the same page.
The database will be enriched, updated, and maintained continually.

\subsection{\gls{bn} training process using artificially generated \emph{train} and \emph{validation} datasets}

In order to train and validate the \gls{bn}, the artificially generated data introduced in section \textbf{Data preparation} were used.
Afterwards, the \gls{bn} was evaluated on both artificial and experimental data also introduced in section \textbf{Data preparation}.
The traces in each dataset were segmented in temporal windows of 0.5 s.
In regards to the artificially generated datasets, we ended up with 60,000 temporal windows for training, 30,000 for validation and 30,000 for testing.
Regarding experimental real datasets we ended up with 852 temporal windows from our $\lambda$-DNA experimental dataset and 1,512 windows from our \emph{streptavidin} experimental dataset.

The \gls{bn} is composed of two networks--\gls{resnet} 1 and \gls{resnet} 2, which were trained separately.
Five different \gls{bn} instances were trained using 5 different \glspl{snr} in the artificial datasets (\gls{snr}=4, 2, 1, 0.5 and 0.25).
In all cases, smooth $l_1$-loss and \gls{sgd} optimisation were adopted. Smooth $l_1$-loss can be seen as a blend of $l_1$-loss and $l_2$-loss. It returns the $l_1$-loss when the absolute value of the argument is high and the $l_2$-loss when such argument is close to zero. In Eq. \ref{loss}, the quadratic segment smooths the $l_1$-loss near $|x - y| = 0$. This loss is less sensitive to \emph{outliers} than other measures such as the \gls{mse} loss. In some cases, it could prevent exploding gradients, which is desirable in networks with architectures as the one used in this work \cite{girshick2015fast} .

\begin{equation}
    \label{loss}
    loss =
    \begin{cases}
    0.5(x-y)^2, & \text{if } |x-y| < 1\\
    |x-y|-0.5, & \text{otherwise}
    \end{cases}
\end{equation}

Smooth $l_1$-loss combines the advantages of $l_1$-loss (producing steady gradients for large values of $|x-y|$) and $l_2$-loss (producing less oscillations during updates when $|x-y|$ is small).

For \gls{snr} equal to 4, 2 and 1, a batch size of 32 temporal windows and an initial learning rate of 0.001 were used. The learning rate was decreased by 90\% every 10 epochs for \gls{resnet} 1 and every 20 epochs for \gls{resnet} 2. For \gls{snr} equal to 0.5 and 0.25, a batch size of 8 windows and an initial learning rate of 0.001 were used. The learning rate was decreased by 20\% every 10 epochs for \gls{resnet} 1 and every 20 epochs for \gls{resnet} 2.
During the \gls{resnet} 2 training procedure, temporal windows without translocation events were ignored. To predict average translocation amplitude and duration, \gls{resnet} 2 always received windows containing at least one translocation. On the other hand, \gls{resnet} 1 processed all temporal windows during training--with and without translocation events.
During training, \gls{resnet} 2 did not receive the prediction of the number of pulses in the window from \gls{resnet} 1, it instead received such information from the ground-truth in the dataset.
Such information was injected in \gls{resnet} 2 in its \gls{ffnn}.

Validation after each epoch in the training process was implemented on each network.
For \gls{resnet} 2, the network with the minimum relative errors of amplitude and duration \emph{won} and was saved in each epoch iteration.
The relative error is defined as:

\begin{equation}
    \label{Rerr}
    Err = \frac{|x - x_0|}{x_0}100\%,
\end{equation}

where, $x$ is the predicted value and $x_0$ the true value.
\gls{resnet} 1, on the other hand, may process temporal windows without translocation spikes.
Consequently it has a division-by-zero risk if the relative error in Eq. \ref{Rerr} is applied.
Therefore, for the \gls{resnet} 1 validation, \gls{rpd}, defined by Eq. \ref{rpd}, was adopted to represent the relative error of predicted pulse number.

\begin{equation}
   \label{rpd}
    RPD = \frac{|x - x_0|}{(|x|+|x_0|)/2}
\end{equation}

\subsection{Evaluation (testing) process using artificially generated \emph{test} dataset}

In order to evaluate the \gls{bn}, a held-out artificially generated dataset, used for neither training nor validation, was used.
This dataset has the same size as the validation dataset. Since it is artificially generated, it is completely labelled.
Therefore, for each temporal window from the traces, the ground-truth features were known during data generation, including the real number of pulses and the real average amplitude and duration of the pulses in the temporal windows.
Then, the relative error between the labels and the predictions produced by the \gls{bn} was calculated, following Eq. \ref{Rerr}.

Unlike training and validation, during evaluation, the \gls{bn} works as follows:
First, \gls{resnet} 1 processes the temporal window and outputs an estimation of the number of pulses in the window.
If the number of pulses predicted by \gls{resnet} 1 is 0, \gls{resnet} 2 will not process the input and the \gls{bn} will predict 0 for pulses, 0 average amplitude and 0 average duration.
On the other hand, if \gls{resnet} 1 predicts one or more pulses in the window, then \gls{resnet} 2 will process the input and predict two aforementioned features, the average amplitude and duration of the pulses in the window.
It is important to highlight that \gls{resnet} 2 was trained using only temporal windows with one or more translocation events (pulses). It never received windows without pulses, consequently, and only in case of correct prediction of number of pulses, \gls{resnet} 1 prevented \gls{resnet} 2 from processing windows without pulses, for which it was not trained.

In addition, considering the division-by-zero risk, those two rules below were followed for the evaluation stage: 

\begin{itemize}
    \item If \gls{resnet} 1 correctly predicts 0 pulses in a window, then we consider 0\% error for all features--number of pulses, and average amplitude and duration.
    \item If \gls{resnet} 1 erroneously predicts more than 0 pulses in a window when the ground-truth is actually 0, then we consider 100\% error for all features--number of pulses, and average amplitude and duration.
\end{itemize}



\section{Author information}

\subsection{Corresponding Author}

*E-mail: shili.zhang@angstrom.uu.se

\subsection{Author Contributions}

D.D.: conceptualization, data curation, neural network development and implementation, formal analysis, investigation, validation, visualization, writing–original draft, and writing (review and editing).  C.W.: conceptualization, data curation, data generation, translocation experiment implementation, formal analysis, investigation, validation, visualization, writing (original draft, review and editing).  M.P.: investigation, discussion, writing (review and editing).  D.Z.: result analysis and discussion, writing (review and editing).  S.-L.Z.: project administration, supervision, resources, investigation, writing (original draft, review and editing).
D.D and C.W. contributed equally.

\subsection{Notes}

The authors declare no competing financial interests.

\begin{acknowledgement}

The authors thank Dr. Shuangshaung Zeng for providing the nanopore chips.  They acknowledge the Graphical Processing Units (GPUs) publicly provided by the Google Collaboratory service for training, validating and evaluating the neural network architecture proposed in this work.  This work was partially financially supported by Stiftelsen Olle Engkvist Byggmastare (No.194-0646).



\end{acknowledgement}

\begin{suppinfo}

Signal processing flow for nanopore sensing; network architecture; physical models and data generation; evaluation (testing) results of artificially generated test dataset for different \gls{snr}; training history using artificially generated train and validation datasets for different \gls{snr}; comparison between the results from our neural network and the traditional algorithm; Translocation features of $\lambda$-DNA and streptavidin extracted by the \gls{bn} (PDF).


\end{suppinfo}

\bibliography{achemso-demo}

\end{document}


\large

\maketitle

\tableofcontents

\appendixpageoff
\appendixtitleoff
\renewcommand{\appendixtocname}{Supplementary material}
\begin{appendices}
\setcounter{chapter}{19}
\crefalias{section}{supp}

\section{Signal processing flow for nanopore sensing}

\begin{figure}[h]
  \centering
  \includegraphics[width=16cm]{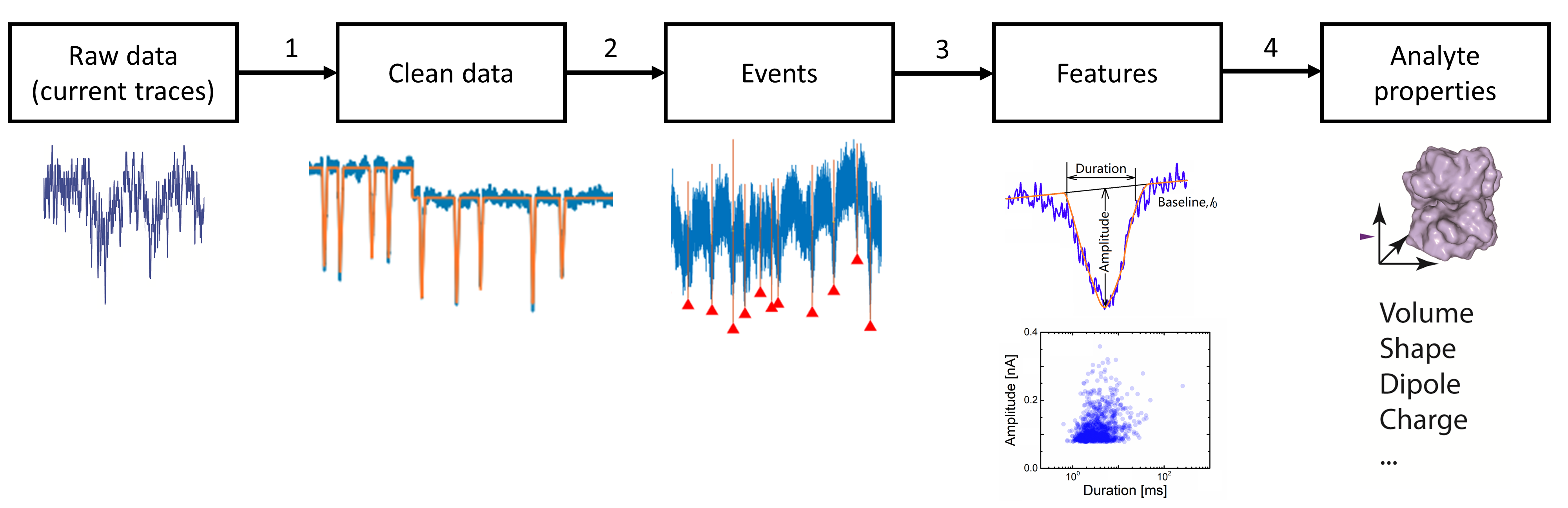}
  \caption{Data processing flow of the nanopore translocation signal.}
  \label{fig:Pross_flow}
\end{figure}

The typical signal processing flow for nanopore sensors is summarised in \textbf{Fig. \ref{fig:Pross_flow}}.

\begin{itemize}
    \item Step 1. Raw data is denoised to form clean data, which can be achieved using low pass filters in frequency domain \cite{pedone2009data}. In time domain, the baseline and blockage level can be traced and cleaned up from the background noise by averaging with a dynamically adjustable threshold, such as CUSUM algorithm \cite{raillon2012fast}. Algorithms based on other theory, such as estimation theory, \emph{e.g.}, Karman filter \cite{o2012kalman}, and wavelet transform \cite{shekar2019wavelet} are also adopted.
    \item Step 2. Translocation events, represented as \emph{spikes}, are recognised and extracted from the current traces. This procedure is usually based on a user-defined threshold of the amplitude as a criterion to separate a true translocation generated spike from a noise fluctuation \cite{zeng2019rectification}.
    \item Step 3. Features of these spikes are extracted based on physical models, such as ADEPT \cite{balijepalli2014quantifying}, peak analysis algorithms, such as DBC \cite{gu2015accurate}, and algorithms of feature analysis in frequency domain, such as Fourier transform and cepstrum \cite{im2018recognition}.
    \item Step 4. Properties of the translocating analytes are inferred from the extracted features. Based on simple physical models, the amplitude of spikes is correlated to the size and shape of analytes \cite{larkin2014high,sha2018identification}.
    The duration is related to the translocation speed and nanopore-analyte interaction, reflecting the physiochemical properties, such as mass, charge, dipole, and hydrophobicity \cite{tsutsui2019solid, houghtaling2018nanopore}.
    The frequency of spikes concerns the concentration of analytes \cite{houghtaling2018nanopore, lan2011nanoparticle}.
    Furthermore, details of translocation waveform are considered by more sophisticated models \cite{houghtaling2018nanopore, houghtaling2019estimation}, such as that using the fingerprint feature of blockage current distribution to distinguish 10 kinds of proteins \cite{yusko2017real}.
    In this step, \gls{ml}-based classification algorithms are widely adopted to cluster the events and associate them to different analytes, such as support vector machine \cite{im2019single}, \gls{cnn} \cite{misiunas2018quipunet}, logistic classifier \cite{wei2019learning}, and decision tree \cite{arima2020digital, arima2020solid}.
\end{itemize}

\section{Network architecture}

Nowadays, thanks to the advent of the \emph{Representation Learning} theory \cite{10.1109/TPAMI.2013.50}, machines can automatically \emph{discover representations}, which is relevant for the detection of specific features demanded by the network designer. In general, \gls{dl} is known as a set of multi-layer representation-learning methods. Starting from the raw input, these networks transform the representations, one layer at a time by means of simple, but non-linear computations. Representations at higher layers are considered to be more acute or \emph{pertinent} for the features pretended to be extracted from the network.
\gls{dl} has shown outstanding breakthroughs in the last years in areas ranging from \gls{cv} to \gls{nlp}, from science to engineering, from medicine to material and computer sciences, etc  \cite{10.1109/TPAMI.2013.50,lecun2015deeplearning,Schmidhuber_2015}.
By only individualising the correct cost function in a subsystem at the top of the network, errors are then backpropagated after each iteration and the weights of the network are automatically adjusted to do better predictions in response to subsequent inputs in a training dataset \cite{lecun2015deeplearning}.


In this work, we introduce a new network named \gls{bn}.
In the \gls{bn}, a specific architecture called \gls{resnet} is adopted. \gls{resnet} is an artificial \gls{dnn} that uses \glspl{cnn} \footnote{A \gls{cnn} is a highly utilized sub-type of \gls{dnn} mostly applied to analyzing visual imagery.} and skips connections, or shortcuts that jump over some layers \cite{he2015deep, 7780459}.
The main motivation for skipping layers is to avoid the problem of \emph{vanishing and exploding gradients}.
By means of these \emph{skip connections} the network can reuse activations from previous layers until posterior layers learn their weights.
In the worst scenario, layer $l + 1$ is able to receive an intact output from layer $l - 1$ if layer $l$ has not yet learned a proper weigh configuration.
In this way, instead of making layer $l + 1$ receive a potentially harmful representation from an \emph{immature} layer $l$, it can avoid utilisation of layer $l$ and instead receive better information from previous layers in the network until a specific training point at which layer $l$ eventually learns its correct weights. It is worth noting that $l \pm 1$ is used here in a figurative way to favour a clearer explanation.
In real implementations, skipping connections actually jump two or more layers and a more realistic scenario would describe something like $l \pm 2$ or $l \pm 3$.
In this manner \gls{resnet} would never go \emph{off rail} in terms of the correct manifold in its learning space.
Nowadays, \gls{resnet} is considered a classic architecture usually employed as a backbone for many computer vision tasks. This architecture gained its privileged place among other highly effective \gls{dl} architectures winning the ILSVRC 2015 in image classification, detection, and localisation, as well as the MS COCO 2015 detection, and segmentation \cite{7780459}.

In this work, we have implemented the \gls{bn} using an architecture called \gls{resnet} 18.
This is one of the standard architectures utilised in \gls{dl} frameworks such as Pytorch.
PyTorch is an open source \gls{ml} library based on the Torch library. It is essentially developed by \gls{fair} laboratory and is used for \gls{ml} applications such as computer vision and natural language processing.
Currently \gls{resnet} 18 presents the performance with Top-1 and Top-5 error rates of 30.24\% and 10.92\% on imagenet dataset respectively \cite{he2015deep}.

\section{Physical models and data generation}
\subsection{Resistance model and open-pore current}
To determine the baseline of the current trace, \emph{i.e.} open-pore current, our previously established resistance model based on the concept of effective transport length is adopted here \cite{Wen2017Rmodel}. The resistance of a cylindrical nanopore can be expressed as

\begin{equation}
    \label{Res}
    R = \frac{4\rho L_{eff}}{\pi d_p^2}
\end{equation}

where, $\rho$ is the resistivity of the electrolyte, $d_p$ the diameter of the nanopore and $L_{eff}$ the effective transport length of the nanopore that is defined as the sum of the distances from the location inside the nanopore where the electric filed is the highest to the two opposite points along the central axis of the pore where the electric fields both fall to $e^{-1}$ of the maximum. For a cylinder pore, $L_{eff}$ is \cite{Wen2017Rmodel}

\begin{equation}
    \label{Leff}
    L_{eff} = 0.92d_p+h
\end{equation}

where, $h$ is the thickness of nanopore. The resistivity of the electrolyte is determined by its salt concentration, $c_0$ and the mobility of cations, $\mu_c$ and anions $\mu_a$.

\begin{equation}
    \label{rou}
    \rho = (q N_A c_0 (\mu_c + \mu_a))^{-1}
\end{equation}

where, $q$ is the element charge and $N_A$ the Avogadro constant.
The surface charge on the nanopore sidewall also contributes to the conductance. The surface conductance can be expressed as

\begin{equation}
    \label{Gs}
    G_s = \mu \sigma \frac{\pi d_p}{h}
\end{equation}

where, $\mu$ is the mobility of the counterions in the surface electric double layer and $\sigma$ the surface charge density. Thus, at a given bias voltage $V$, the open-pore current is

\begin{equation}
    \label{I0}
    I_0 = V (G_s+1/R)
\end{equation}

Values of the parameters used in this model are listed in Table  \ref{tab:tab1}.

\begin{table}[h]
\centering
\caption{Parameters used in the resistance model}\label{tab:tab1}
$\begin{tabu}{|l|l|l|l|}
\hline
\text{Parameter} & \text{Unit} & \text{Value} & \text{Reference} \\\hline
q & C & 1.6 \times 10^{-19} & \\\hline
N_A & mol^{-1} & 6.02 \times 10^{23} & \\\hline
c_0 & mM & 100 & \\\hline
d_p & nm & 20 & \\\hline
h & nm & 20 & \\\hline
V & mV & 300 & \\\hline
\sigma & Cm^{-2} & -0.02 & \cite{anderson2013ph,kox2010local,luan2013electro} \\\hline
\mu_c & m^2V^{-1}s^{-1} & 7.575 \times 10^{-9} & \cite{adamson2012textbook} \\\hline
\mu_a & m^2V^{-1}s^{-1} & 7.874 \times 10^{-9} & \cite{adamson2012textbook} \\\hline
\end{tabu}$
\end{table}

\subsection{Translocation model and translocation spikes}
\begin{figure}[H]
  \centering
  \includegraphics[width=6cm]{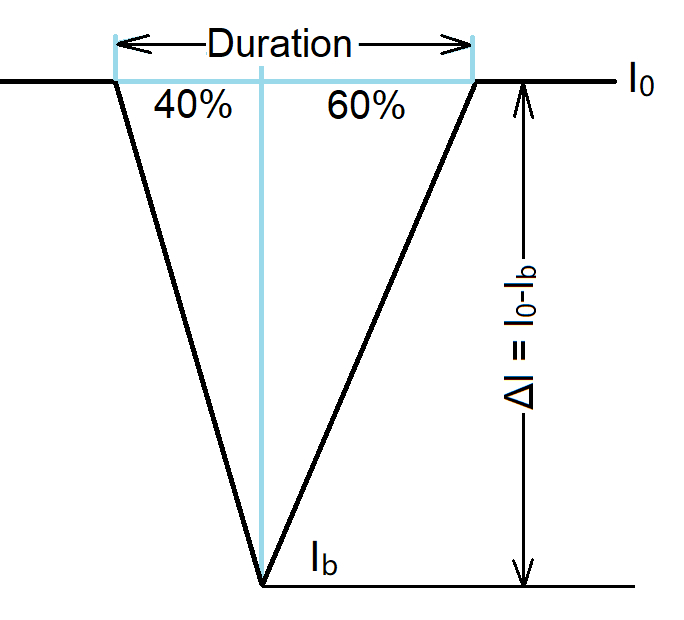}
  \caption{The triangular spike waveform used in signal generation}
  \label{fig:waveform}
\end{figure}

The amplitude of the translocation spikes are simply determined by a steric blockage model \cite{Wen2016DNAsequencing}, that concerns the ratio of the cross-section area of the nanopore to that of the translocating nanosphere.

\begin{equation}
    \label{Ib}
    \Delta I = I_0 \frac{D_{np}^2}{d_p^2}
\end{equation}

where, \gls{dnp} is the diameter of the translocating nanosphere. The shape of translocation spikes is approximated by a triangle, as shown in\textbf{Fig. \ref{fig:waveform}}. In a spike, the current decreases from the open-pore current $I_0$ in the first 40\% of the duration time to reach the minimum $I_b$, and then increases back to $I_0$ in the rest of the 60\% duration.

The probability of finding a spike at certain time point is set to be proportional to the nanosphere concentration \gls{cnp} and exponentially dependent on bias voltage.

\begin{equation}
    \label{Prob}
    P = k_0 C_{np} e^{\frac{qV}{k_B T}}
\end{equation}

where, $k_0$ is a coefficient, $k_B$ the Boltzmann constant and $T$ temperature in Kelvin. During the signal generation, a time step $\Delta t$ of 0.1 ms, \emph{i.e.} sampling rate of 10 kHz, is specified. Then, a state of either “open-pore” (o) or “blockage” (b) is randomly generated for each $\Delta t$ according to a two-point distribution based on probability $P$. Each pore is sequentially accessed during one $\Delta t$ and $I_0$ is assigned for pores in state “o” and $I_b$ for pores in state “b”. It is worth to mention that a pore cannot change to state “o” until lasting for a certain duration time, once it is in state “b”. The same algorithm of signal generation and the same dependence of spike appearance probability on analyte concentration and bias voltage have been adopted in our previous work for multiple nanopores \cite{Wen2018group}. Values of the parameters used in this model are listed in Table \ref{tab:tab2}.

\begin{table}[h]
\centering
\caption{Parameters used for translocation spike generation}\label{tab:tab2}
$\begin{tabu}{|l|l|l|l|}
\hline
\text{Parameter} & \text{Unit} & \text{Value} & \text{Reference} \\\hline
k & nM^{-1} & 9.1 \times 10^{-8} & \cite{Wen2018group} \\\hline
k_B & JK^{-1} & 1.38 \times 10^{-23} & \\\hline
T & K & 300 & \\\hline
C_{np} & nM & 0.01-1 & \cite{Wen2018group} \\\hline
Duration & ms & 0.5-5 & \\\hline
\end{tabu}$
\end{table}

\subsection{Noise model and background noise generation}
A comprehensive noise model of solid-state nanopore has been established based on experimental characterisation of $SiN_X$ nanopores \cite{Wen2017noise}. There are four distinct noise sources in the frequency range below 5 kHz: flicker noise $S_{IF}$, electrode noise $S_{IE}$, white thermal noise $S_{IT}$, and dielectric noise $S_{ID}$. The \gls{psd} of these noise components are:

\begin{equation}
    \label{SIF}
    S_{IF} = \frac{\alpha_H I_0^2}{N_c f^{\beta_1}}
\end{equation}

\begin{equation}
    \label{SIE}
    S_{IE} = \frac{\alpha_e}{f^{\beta_2}}
\end{equation}

\begin{equation}
    \label{SIT}
    S_{IT} = \frac{4 k_B T}{R}
\end{equation}

\begin{equation}
    \label{SID}
    S_{ID} = 8 \pi k_B T d_L C_{chip} f
\end{equation}

with

\begin{equation}
    \label{NC}
    N_c = \frac{1}{2} \pi d_p^2 h c_0 N_A
\end{equation}

where, $\alpha_H$ is a constant named Hooge's parameter, $N_c$ the total number of conducting carriers in the nanopore, $f$ the frequency, $\alpha_e$ the current noise parameter for the electrodes, $\beta_i$ the factor for frequency dependency, $d_L$ dielectric loss factor of the nanopore membrane and $C_{chip}$ the parasitic capacitance of the membrane. The total \gls{psd} of the background current noise is the sum of these four components.

\begin{equation}
    \label{psd_total}
    S_I = S_{IF}+S_{IE}+S_{IT}+S_{ID}
\end{equation}

The time sequence of noise with the required length in time domain can be generated based on the white Gaussian noise generator in the MATLAB function library. Using the Fast Fourier Transform and Inverse Fourier Transform, the white Gaussian noise source can be modulated by the total \gls{psd} to become the coloured Gaussian noise. Details of the algorithm can be found in \cite{Wen2016DNAsequencing}. The amplitude of the noise can be tuned by a factor, so does \gls{snr}. Values of the parameters used in the noise model are listed in Table \ref{tab:tab3}.

\begin{table}[h]
\centering
\caption{Values of parameters used in the noise model}\label{tab:tab3}
$\begin{tabu}{|l|l|l|l|}
\hline
\text{Parameter} & \text{Unit} & \text{Value} & \text{Reference} \\\hline
\alpha_H & 1 & 1.9 \times 10^{-4} & \cite{smeets2009low, smeets2008noise} \\\hline
\beta_1 & 1 & 1 & \cite{Wen2017noise} \\\hline
\beta_2 & 1 & 1.5 & \cite{Wen2017noise} \\\hline
\alpha_e & A^2 & 2 \times 10^{-24} & \cite{Wen2017noise, Wen2020review}\\\hline
d_L & 1 & 0.27 & \cite{Wen2017noise}\\\hline
C_{chip} & nF & 52 & \cite{Wen2017noise}\\\hline
\end{tabu}$
\end{table}

\subsection{Baseline variation}
Two kinds of baseline variations are involved in the signal generation: sudden jump of the baseline and slow fluctuation. The sudden jump of the baseline is achieved by randomly appeared steps on the baseline. The height of these steps is set to be 30\% of $\Delta I$ with a random fluctuation of 10\%. The number of steps appeared in a 10 s period is also a randomly generated number with a expectation of 30.
The slow fluctuation of baseline is represented by the superposition of 8 terms of sine and cosine functions.

 \begin{equation}
    \label{baseline_fluc}
    \begin{split}
        I_{fluc} = a_0 I_0 &(a_1 sin(\omega t)+a_2 sin(2\omega t)+a_3 sin(3\omega t)+a_4 sin(4\omega t)+\\
        &b_1 cos(\omega t)+b_2 cos(2\omega t)+b_3 cos(3\omega t)+b_4 cos(4\omega t))
    \end{split}
\end{equation}

The general amplitude of the slow fluctuation is controlled by the factor $a_0$, which is 0.003 in the signal generation. $a_i$ and $b_i$ are amplitude coefficients.They are random numbers with an expectation of zero and \gls{std} showing in Table \ref{tab:tab4}. 

\begin{table}[h]
\centering
\caption{Parameters used in the baseline variations}\label{tab:tab4}
\begin{tabular}{|c|c|c|c|}
\hline
Parameter & Fluctuation & Parameter & Fluctuation \\\hline
$a_1$ & 0.5 & $a_2$ & 0.5\\\hline
$a_3$ & 0.1 & $a_4$ & 0.05\\\hline
$b_1$ & 0.5 & $b_2$ & 0.5\\\hline
$b_3$ & 0.1 & $b_4$ & 0.05\\\hline
\end{tabular}
\end{table}

\newpage

\section{\gls{bn} evaluation (testing) results when processing artificially generated \emph{test} dataset for different \gls{snr}}

\begin{figure}[H]
  \centering
  \includegraphics[width=5cm]{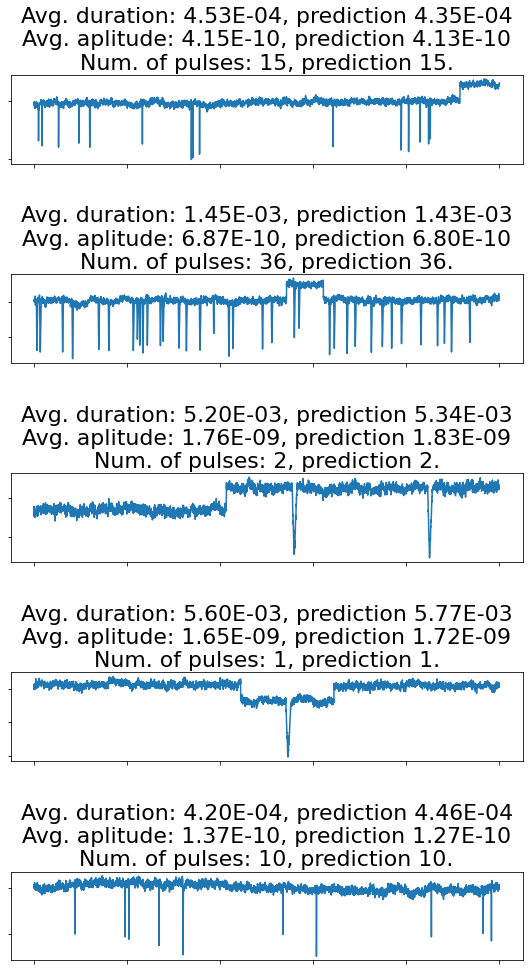}
  \caption{Examples of how the \gls{bn} predicts features from temporal windows with \gls{snr}=4.
           In this batch with 5 temporal windows, the
           average translocation duration error is 3.4\%,
           average translocation amplitude error is 3.3\%,
           average translocation counter error is 0.0\% and
           0 improper measures.
           Improper measures are produced when the ground-truth establishes 0 number of pulses but the network predicts one or more pulses.}
  \label{fig:Examples_SNR_4}
\end{figure}

\begin{figure}[H]
     \centering
     \begin{subfigure}[b]{0.72\textwidth}
         \centering
         \includegraphics[width=\textwidth]{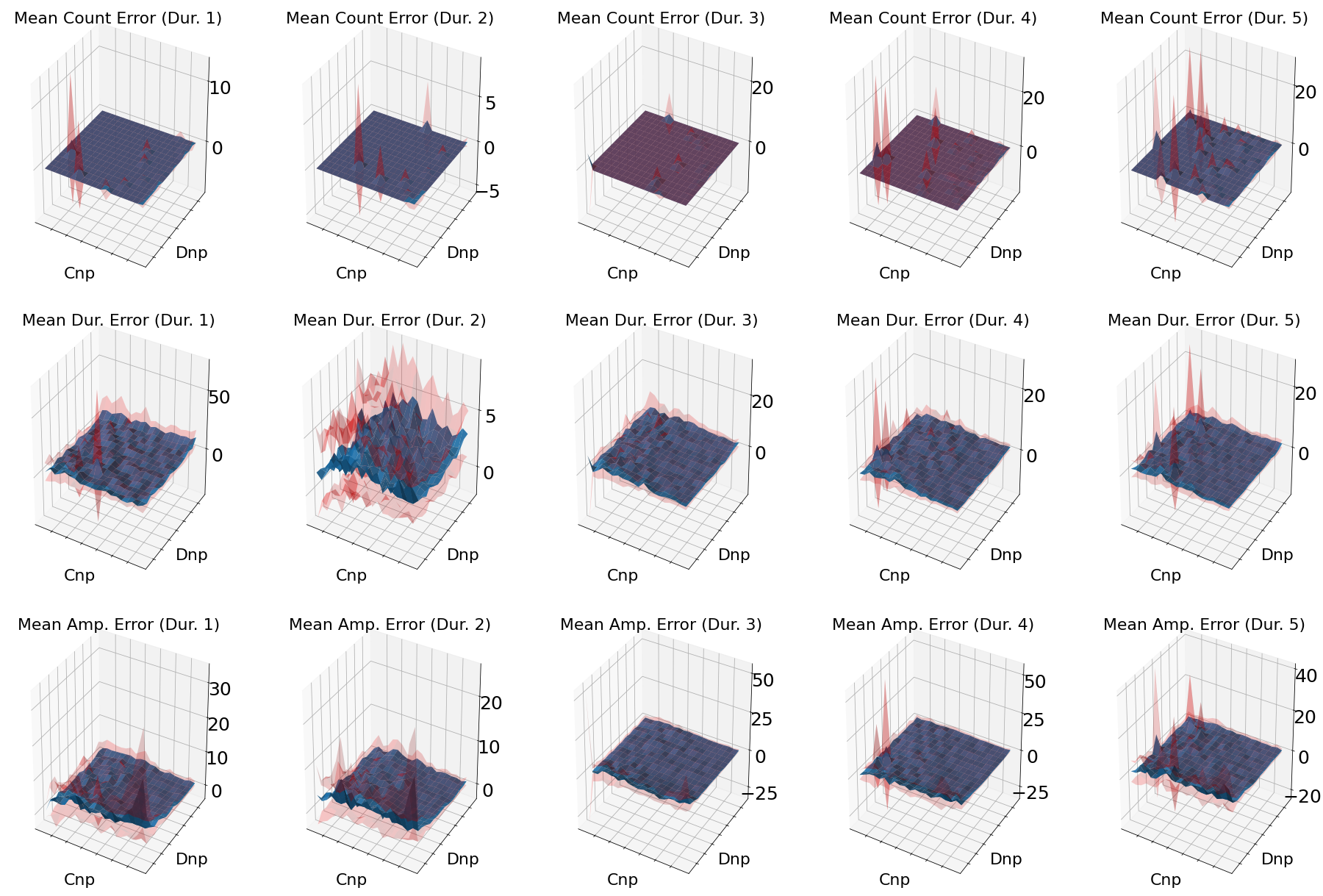}
         \caption{Surfaces of average prediction errors for the different features on temporal windows for different durations.
                  Each surface is distributed through the space of \gls{cnp} and \gls{dnp} values.}
         \label{fig:Surfaces_SNR_4}
     \end{subfigure}
     \hfill
     \begin{subfigure}[b]{0.27\textwidth}
         \centering
         \includegraphics[width=\textwidth]{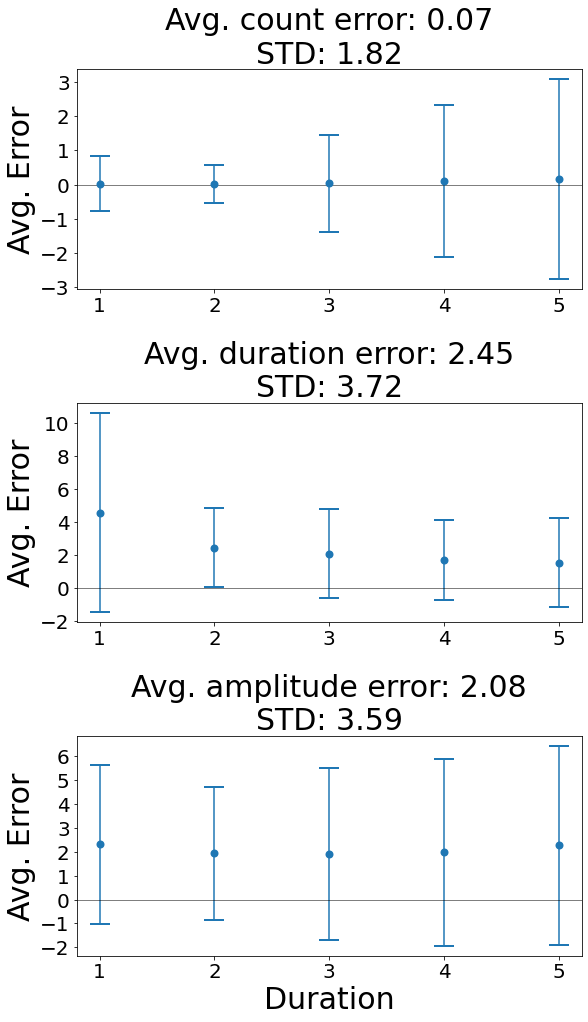}
         \caption{Average prediction errors for each duration. Error bars correspond to \glspl{std}.}
         \label{fig:Bars_SNR_4}
     \end{subfigure}
        \caption{\gls{bn} evaluation on the \gls{snr}=4 dataset.}
        \label{fig:bn_eva_snr_4}
\end{figure}

\begin{figure}[H]
  \centering
  \includegraphics[width=5cm]{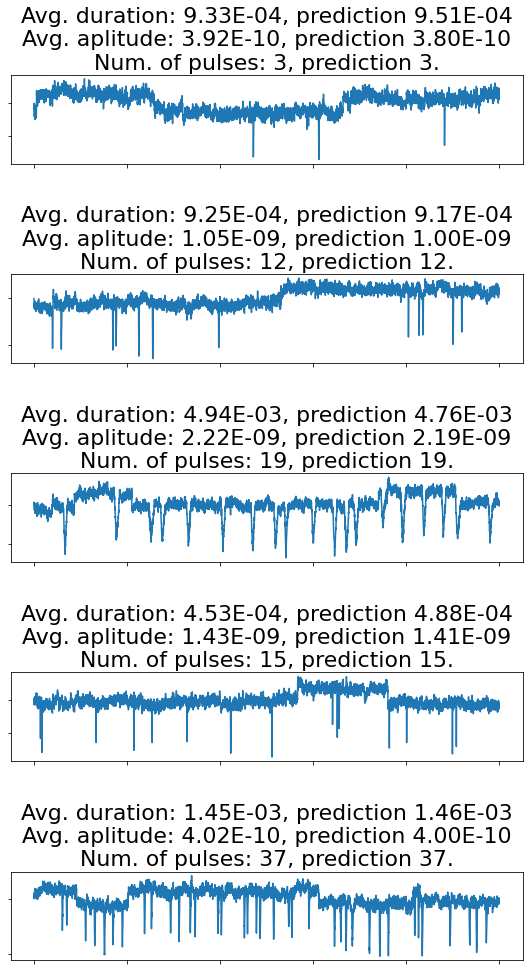}
  \caption{Examples of how the \gls{bn} predicts features from temporal windows with \gls{snr}=2.
           In this batch with 5 temporal, the
           average translocation duration error is 2.8\%,
           average translocation amplitude error is 2.0\%,
           average translocation counter error is 0.0\% and
           0 improper measures.}
  \label{fig:Examples_SNR_2}
\end{figure}

\begin{figure}[H]
     \centering
     \begin{subfigure}[b]{0.72\textwidth}
         \centering
         \includegraphics[width=\textwidth]{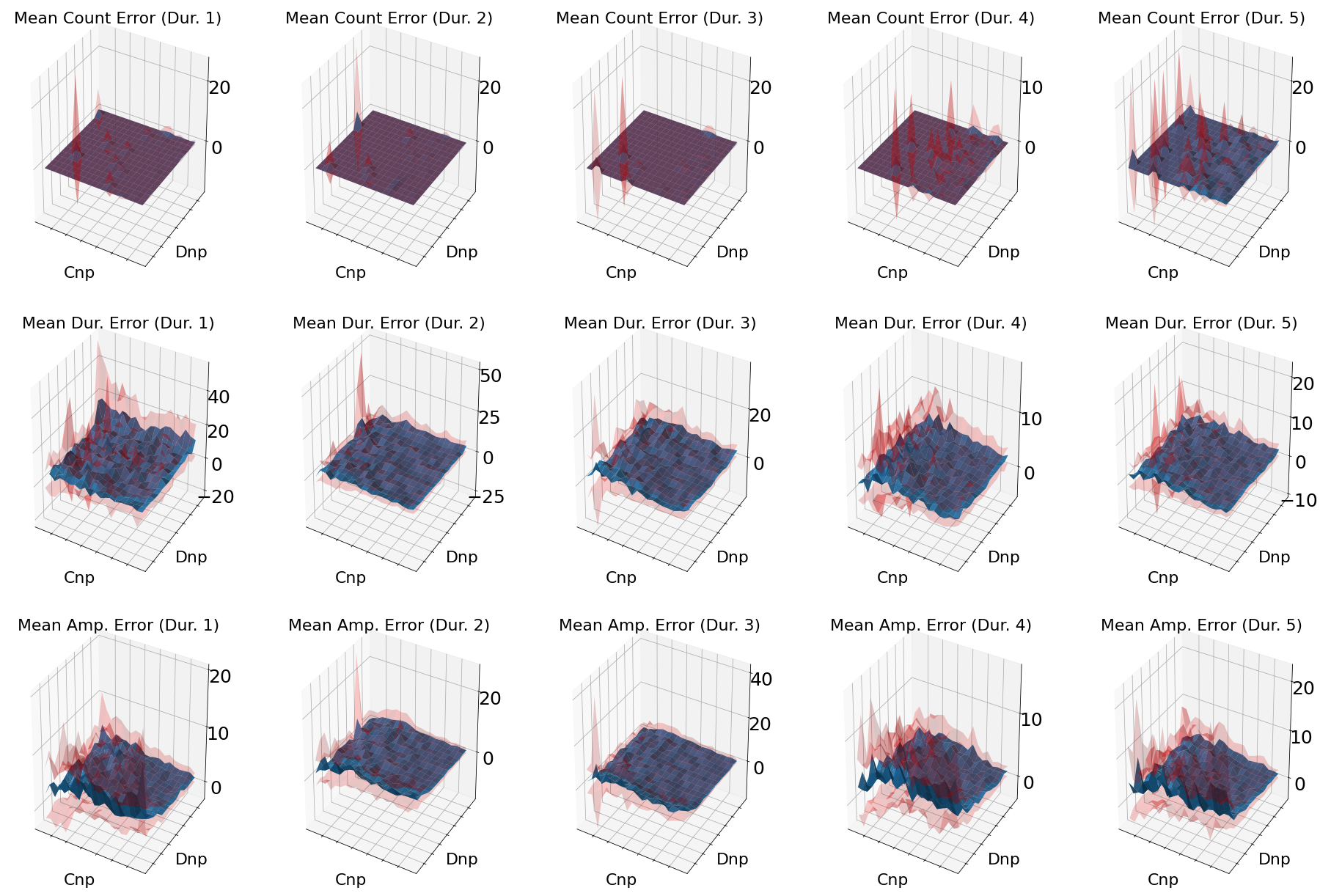}
         \caption{Surfaces of average prediction errors for the different features on temporal windows for different durations.
                  Each surface is distributed through the space of \gls{cnp} and \gls{dnp} values.}
         \label{fig:Surfaces_SNR_2}
     \end{subfigure}
     \hfill
     \begin{subfigure}[b]{0.27\textwidth}
         \centering
         \includegraphics[width=\textwidth]{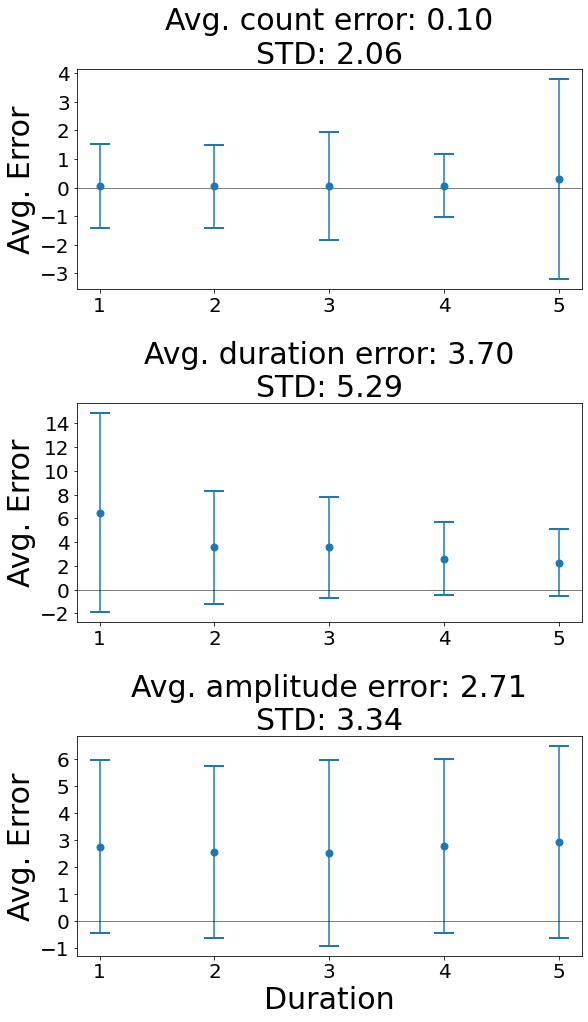}
         \caption{Average prediction errors for each duration. Error bars correspond to \glspl{std}.}
         \label{fig:Bars_SNR_2}
     \end{subfigure}
        \caption{\gls{bn} evaluation on the \gls{snr}=2 dataset.}
        \label{fig:bn_eva_snr_2}
\end{figure}

\begin{figure}[H]
  \centering
  \includegraphics[width=5cm]{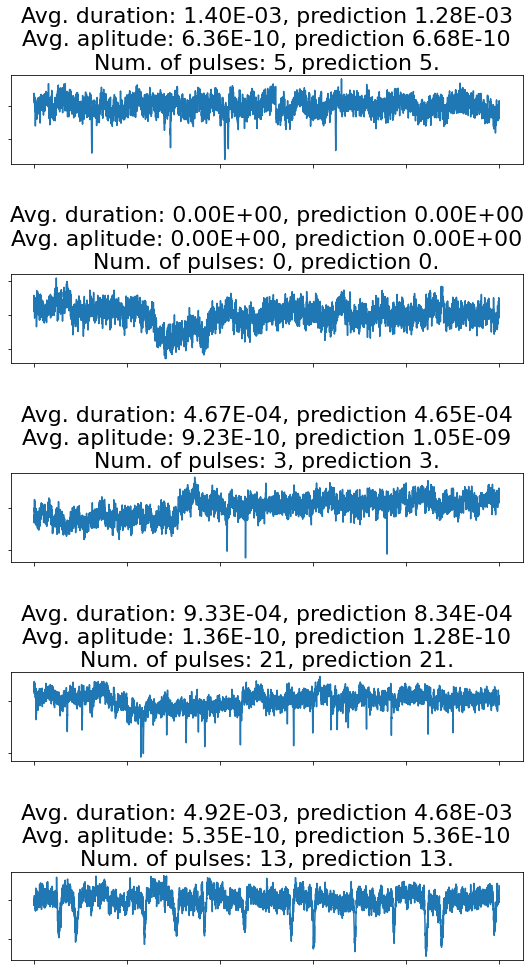}
  \caption{Examples of how the \gls{bn} predicts features from temporal windows with \gls{snr}=1.
           In this batch with 5 temporal windows, the
           average translocation duration error: 5.0\%,
           average translocation amplitude error: 4.8\%,
           average translocation counter error: 0.0\% and
           0 improper measures.}
  \label{fig:Examples_SNR_1}
\end{figure}

\begin{figure}[H]
     \centering
     \begin{subfigure}[b]{0.72\textwidth}
         \centering
         \includegraphics[width=\textwidth]{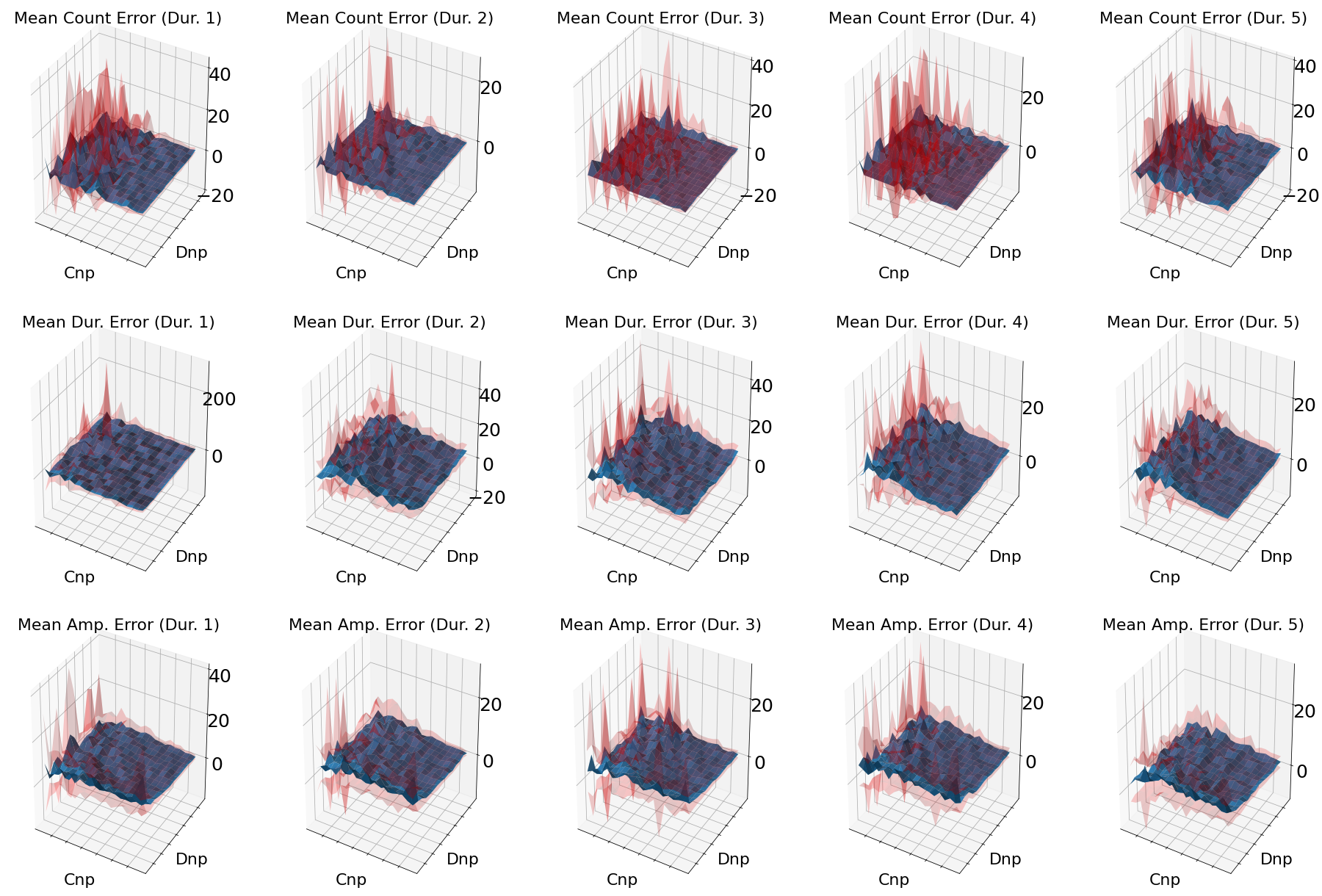}
         \caption{Surfaces of average prediction errors for the different features on temporal windows for different durations.
                  Each surface is distributed through the space of \gls{cnp} and \gls{dnp} values.}
         \label{fig:Surfaces_SNR_1}
     \end{subfigure}
     \hfill
     \begin{subfigure}[b]{0.27\textwidth}
         \centering
         \includegraphics[width=\textwidth]{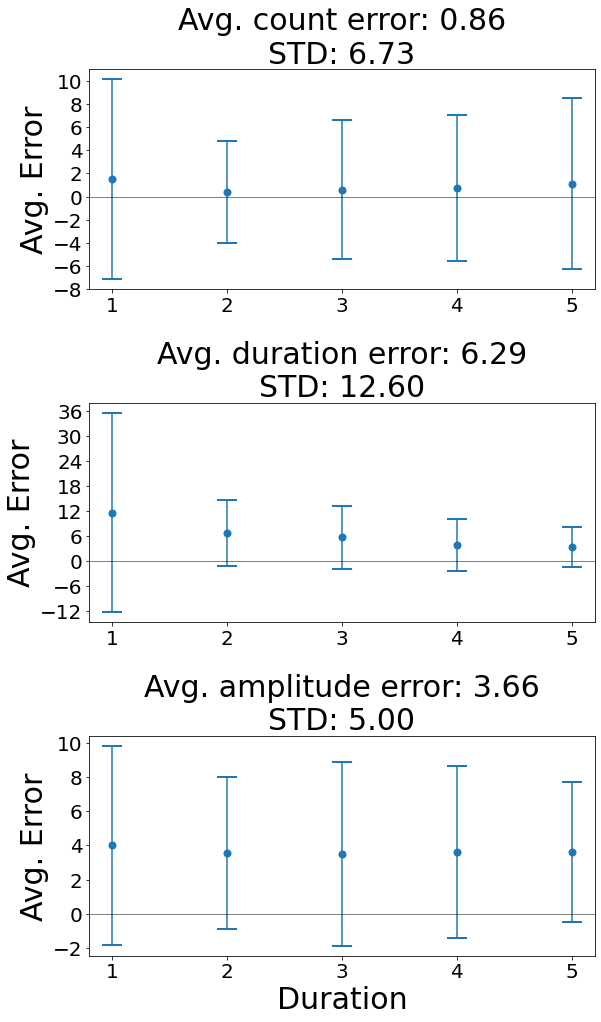}
         \caption{Average prediction errors for each duration. Error bars correspond to \glspl{std}.}
         \label{fig:Bars_SNR_1}
     \end{subfigure}
        \caption{\gls{bn} evaluation on the \gls{snr}=1 dataset.}
        \label{fig:bn_eva_snr_1}
\end{figure}

\begin{figure}[H]
  \centering
  \includegraphics[width=5cm]{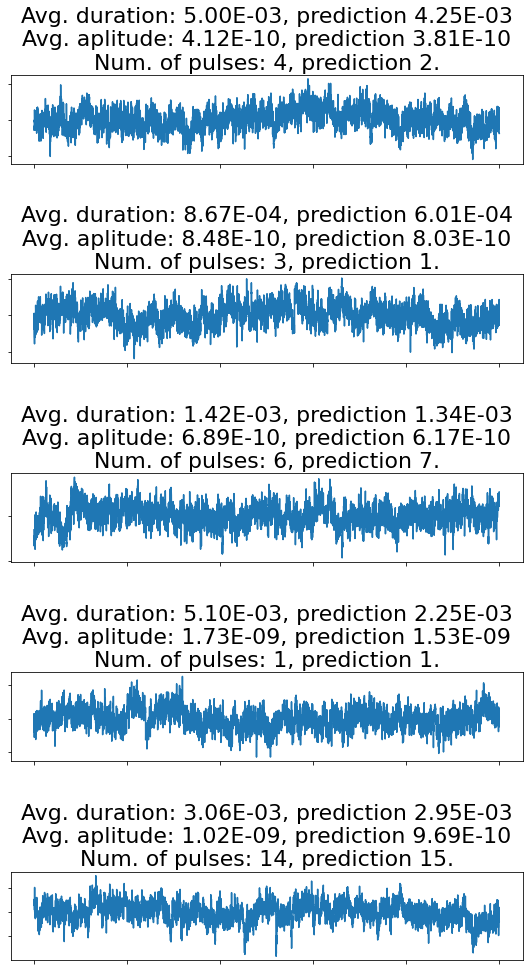}
  \caption{Examples of how the \gls{bn} predicts features from temporal windows with \gls{snr}=0.5.
           In this batch with 5 temporal windows, the
           average translocation duration error: 22.1\%,
           average translocation amplitude error: 8.1\%,
           average translocation counter error: 28.1\% and
           0 improper measures.}
  \label{fig:Examples_SNR_0_5}
\end{figure}

\begin{figure}[H]
     \centering
     \begin{subfigure}[b]{0.72\textwidth}
         \centering
         \includegraphics[width=\textwidth]{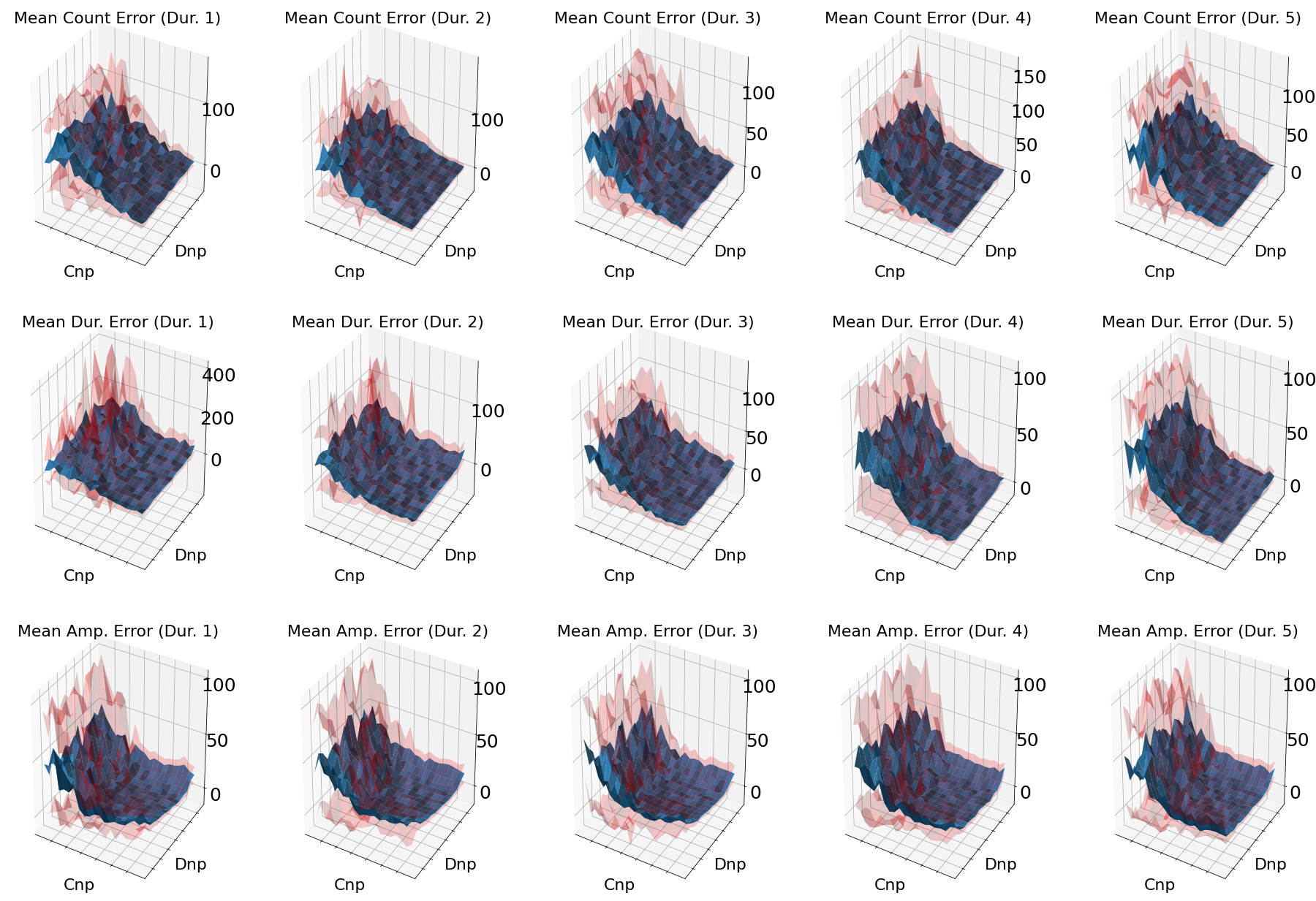}
         \caption{Surfaces of average prediction errors for the different features on temporal windows for different durations.
                  Each surface is distributed through the space of \gls{cnp} and \gls{dnp} values.}
         \label{fig:Surfaces_SNR_0_5}
     \end{subfigure}
     \hfill
     \begin{subfigure}[b]{0.27\textwidth}
         \centering
         \includegraphics[width=\textwidth]{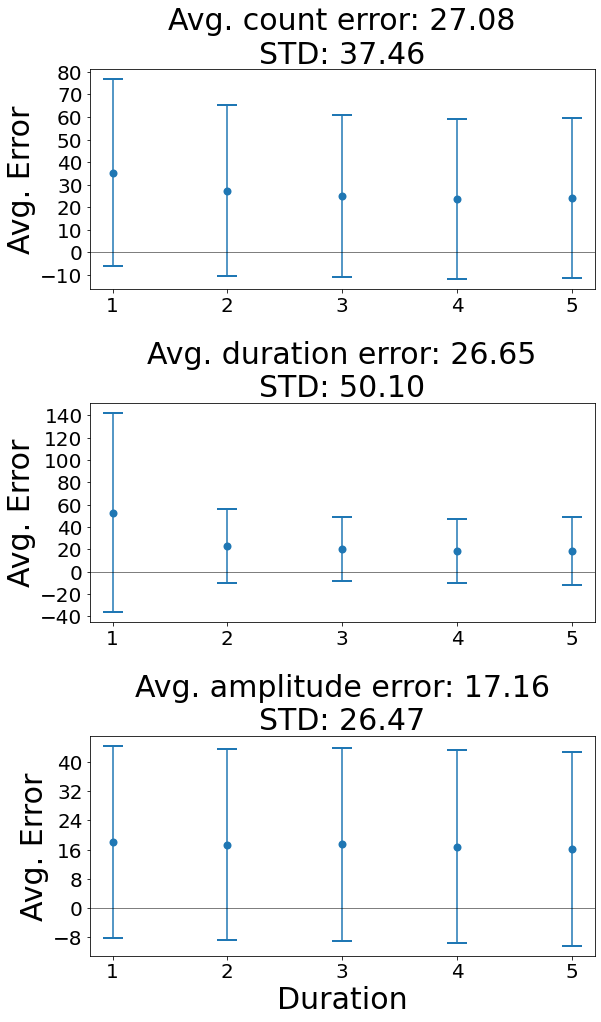}
         \caption{Average prediction errors for each duration. Error bars correspond to \glspl{std}.}
         \label{fig:Bars_SNR_0_5}
     \end{subfigure}
        \caption{\gls{bn} evaluation on the \gls{snr}=0.5 dataset.}
        \label{fig:bn_eva_snr_0_5}
\end{figure}

\begin{figure}[H]
  \centering
  \includegraphics[width=5cm]{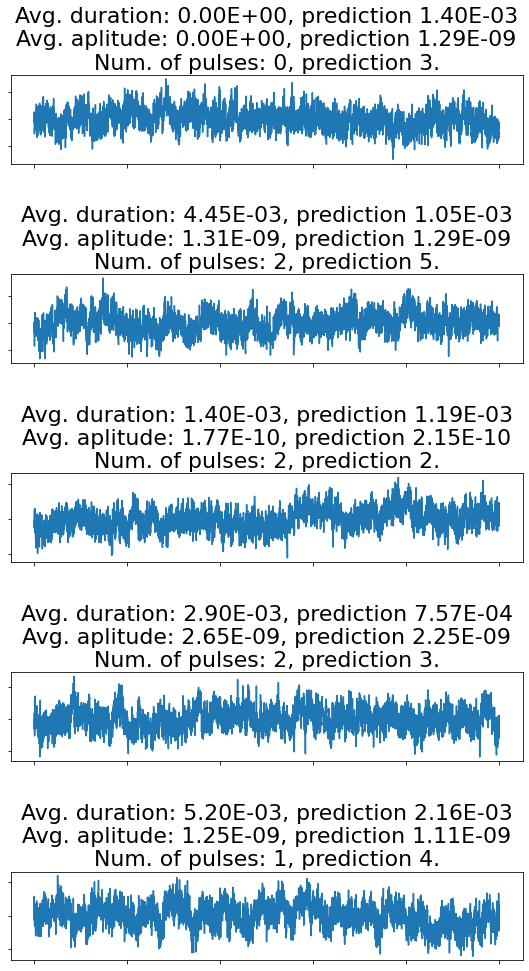}
  \caption{Examples of how the \gls{bn} predicts features from temporal windows with \gls{snr}=0.25.
           In this batch of 5 temporal windows, the
           average translocation duration error: 64.7\%,
           average translocation amplitude error: 30.0\%,
           average translocation counter error: 120.0\% and
           we find 1 improper measure.}
  \label{fig:Examples_SNR_0_25}
\end{figure}

\begin{figure}[H]
     \centering
     \begin{subfigure}[b]{0.72\textwidth}
         \centering
         \includegraphics[width=\textwidth]{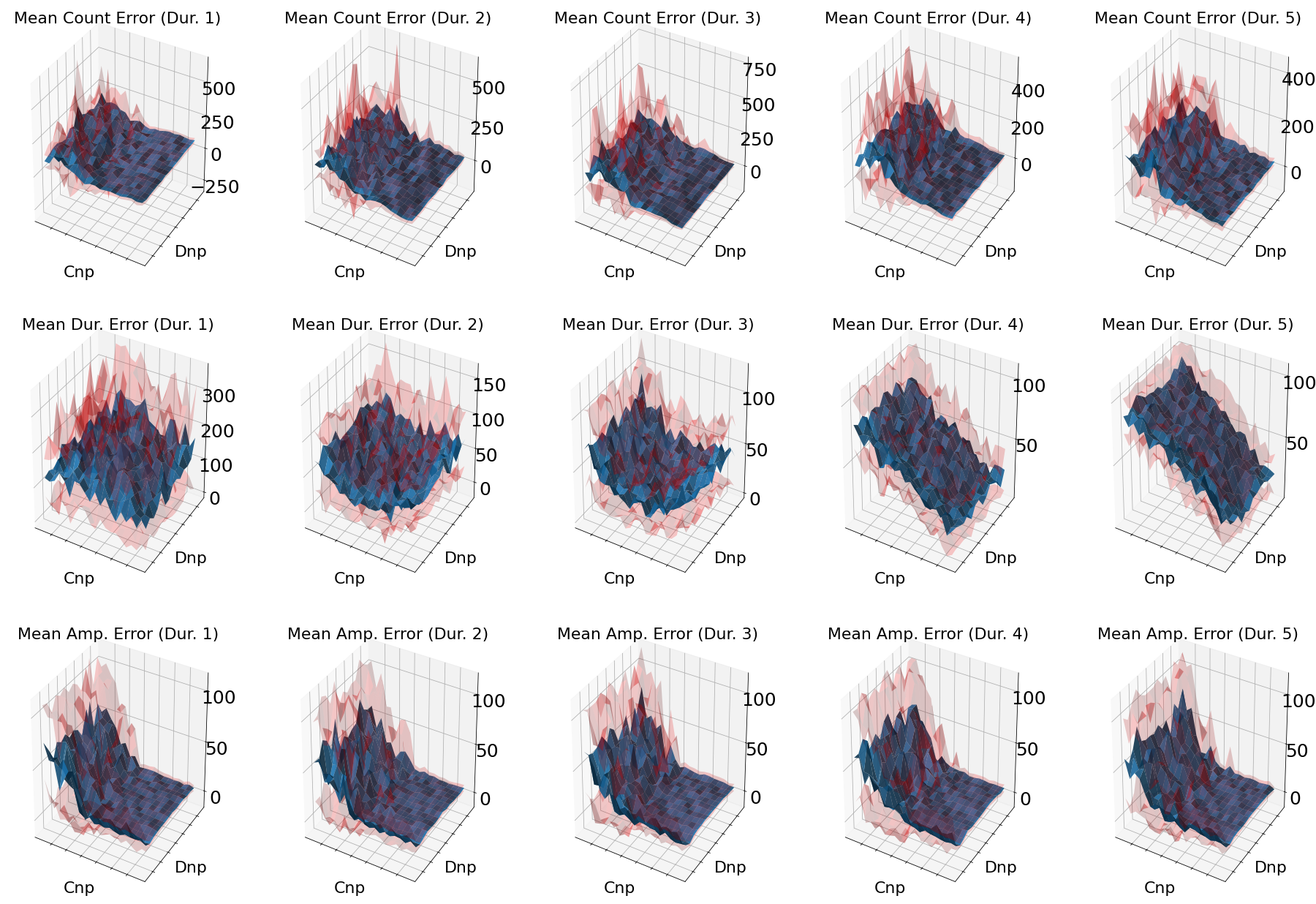}
         \caption{Surfaces of average prediction errors for the different features on temporal windows for different durations.
                  Each surface is distributed through the space of \gls{cnp} and \gls{dnp} values.}
         \label{fig:Surfaces_SNR_0_25}
     \end{subfigure}
     \hfill
     \begin{subfigure}[b]{0.27\textwidth}
         \centering
         \includegraphics[width=\textwidth]{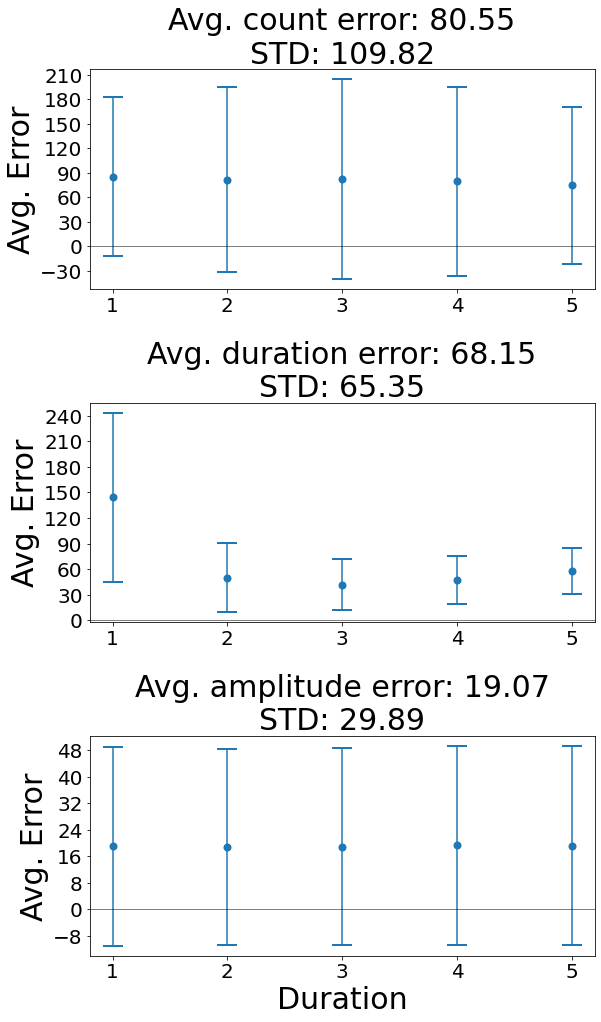}
         \caption{Average prediction errors for each duration. Error bars correspond to \glspl{std}.}
         \label{fig:Bars_SNR_0_25}
     \end{subfigure}
        \caption{\gls{bn} evaluation on the \gls{snr}=0.25 dataset.}
        \label{fig:bn_eva_snr_0_25}
\end{figure}

In \textbf{Figs. \ref{fig:Examples_SNR_4}, \ref{fig:Examples_SNR_2}, \ref{fig:Examples_SNR_1}, \ref{fig:Examples_SNR_0_5}} and \textbf{\ref{fig:Examples_SNR_0_25}}, we can see bathes each with 5 random temporal windows and how the corresponding \gls{bn} instances predict the ground truth features in the windows.
Likewise in \textbf{Figs. \ref{fig:bn_eva_snr_4}, \ref{fig:bn_eva_snr_2}, \ref{fig:bn_eva_snr_1}, \ref{fig:bn_eva_snr_0_5}} and \textbf{\ref{fig:bn_eva_snr_0_25}}, we show the statistical results of \gls{bn} when processing the artificially generated \emph{test} dataset.

\newpage

\gls{dl} has shown advances in noise reduction in general, such as in image noise reduction \cite{liu_overview_2019,tian2020deep,ivanov_deepdenoise_2020,jin_learning_2019}
and audio signals, for speech enhancement \cite{serizel:hal-02506387,kumar2016speech} and its application to cochlear implants \cite{lai_deep_2018}.
From an architectural point of view, it is claimed that \glspl{resnet} perform better than other \emph{simpler} architectures which does not have skip connections.
This hypothesis is based on experiments that suggest that \glspl{resnet} have better noise stability, which is empirically supported for both simplified and fully-fledged \gls{resnet} variations \cite{yu2019identity}.

The relative errors stay near 1\% when \gls{snr} $\geq$ 2, while they can go straight up to 100\% with \gls{snr} reaching 0.25.
For a signal with \gls{snr} $\geq$ 2, the neural network can almost thoroughly recognise every spike from the background noise.
Increasing \gls{snr} does not influence the errors, indicating that these errors come from the system itself, \emph{i.e.} the neural network.
When \gls{snr} is smaller than 1, the errors show a strong dependence on \gls{snr}, indicating that the interference on the spike recognition is from the background noise.
It is almost impossible for traditional algorithms to correctly recognise translocation spikes in a background noise that has a similar amplitude to the spikes, \emph{i.e.} \gls{snr}=1.

\newpage

\section{\gls{bn} training history using artificially generated \emph{train} and \emph{validation} datasets for different \gls{snr}}

\begin{figure}[H]
     \centering
     \begin{subfigure}[b]{0.48\textwidth}
         \centering
         \includegraphics[width=\textwidth]{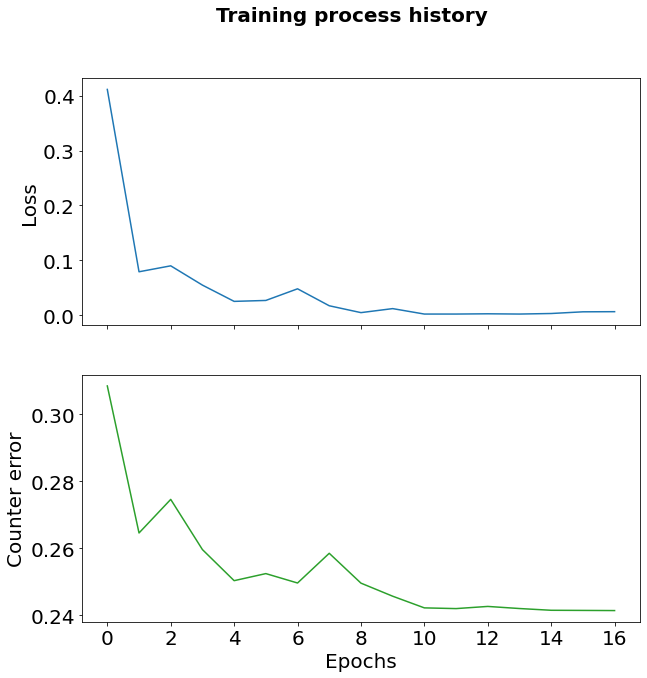}
         \caption{\gls{resnet} 1.
         The total training time was 3 hours, 38 minutes and 50 seconds,
         while the average time during one epoch of training was 0 hours, 12 minutes and 52 seconds.
         Total training took 16 epochs
         The best model was saved in epoch 16.}
         \label{fig:rn1_th_snr4}
     \end{subfigure}
     \hfill
     \begin{subfigure}[b]{0.48\textwidth}
         \centering
         \includegraphics[width=\textwidth]{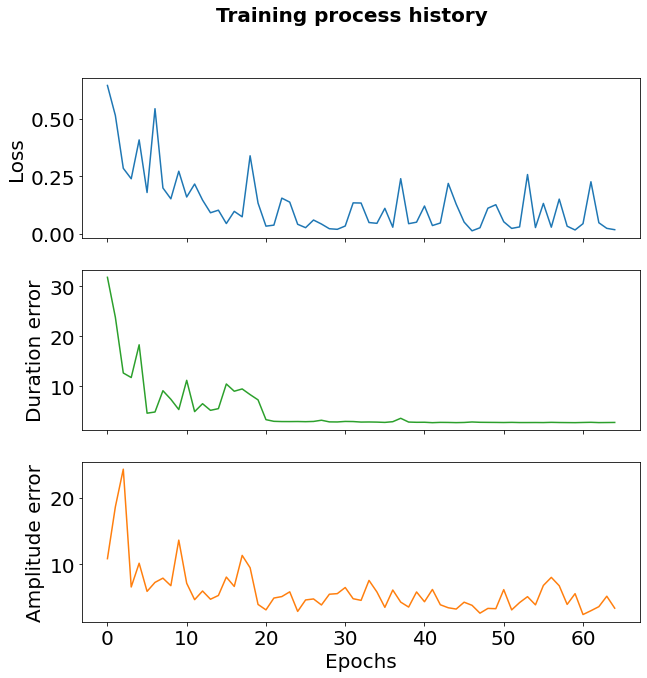}
         \caption{\gls{resnet} 2.
         The total training time was 15 hours, 3 minutes and 47 seconds,
         while the average time during one epoch of training was 0 hours 13 minutes and 54 seconds.
         Total training took 64 epochs.
         The best model was saved in epoch 60 after 14 hours, 10 minutes and 30 seconds.}
         \label{fig:rn2_th_snr4}
     \end{subfigure}
        \caption{Training History trained on the \gls{snr}=4 dataset.}
        \label{fig:bn_th_snr4}
\end{figure}

\begin{figure}[H]
     \centering
     \begin{subfigure}[b]{0.48\textwidth}
         \centering
         \includegraphics[width=\textwidth]{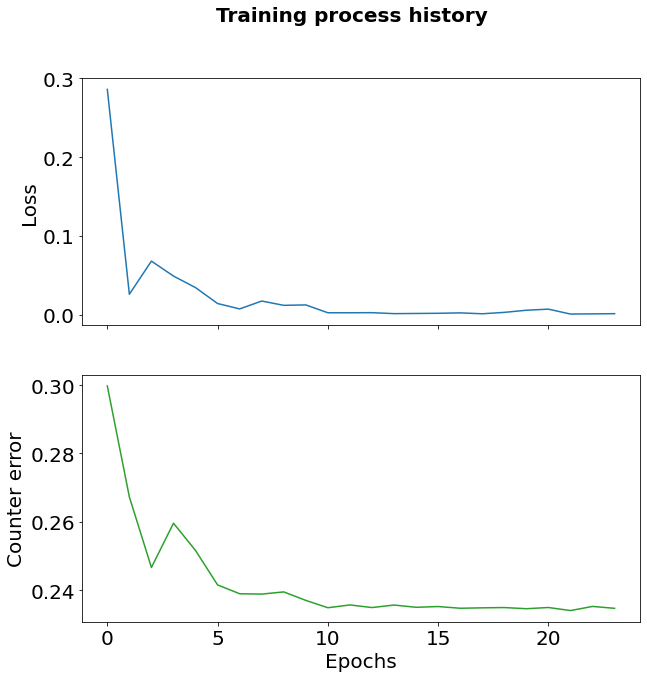}
         \caption{\gls{resnet} 1.
         The total training time was 5 hours, 35 minutes and 55 seconds,
         while the average time during one epoch of training was 0 hours, 13 minutes and 59 seconds.
         Total training took 23 epochs.
         The best model was saved in epoch 21 after 4 hours 55 minutes and 6 seconds.}
         \label{fig:rn1_th_snr2}
     \end{subfigure}
     \hfill
     \begin{subfigure}[b]{0.48\textwidth}
         \centering
         \includegraphics[width=\textwidth]{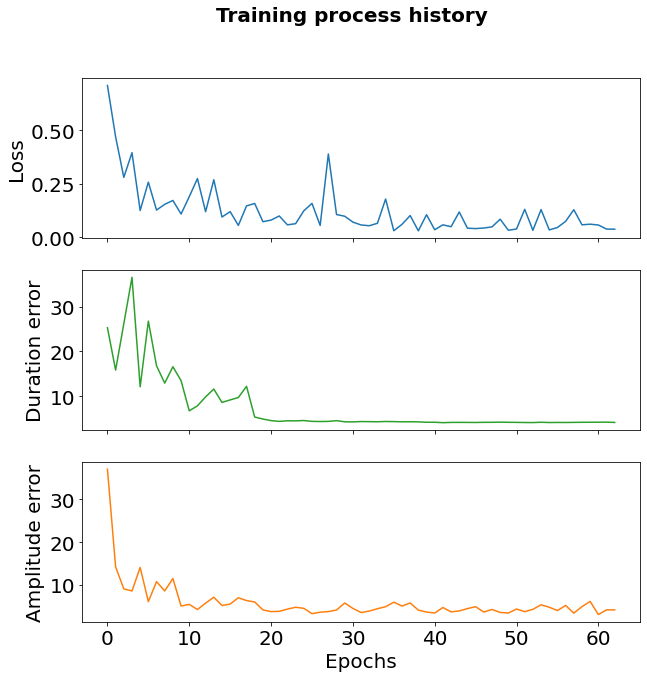}
         \caption{\gls{resnet} 2.
         The total training time was 16 hours, 41 minutes and 36 seconds,
         while the average time during one epoch of training was 0 hours 15 minutes and 53 seconds.
         Total training took 62 epochs.
         The best model was saved in epoch 60 after 16 hours, 13 minutes and 42 seconds.}
         \label{fig:rn2_th_snr2}
     \end{subfigure}
        \caption{Training History trained on the \gls{snr}=2 dataset.}
        \label{fig:bn_th_snr2}
\end{figure}

\begin{figure}[H]
     \centering
     \begin{subfigure}[b]{0.48\textwidth}
         \centering
         \includegraphics[width=\textwidth]{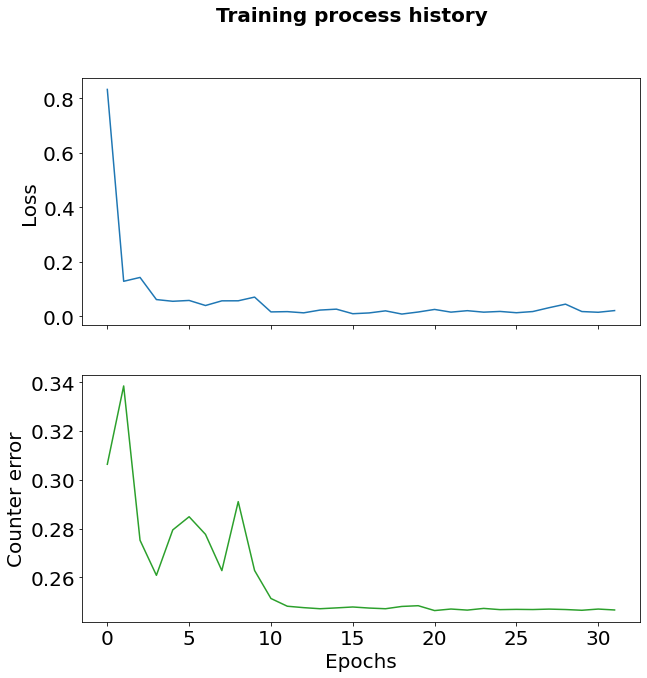}
         \caption{\gls{resnet} 1.
         The total training time was 7 hours, 24 minutes and 29 seconds,
         while the average time during one epoch of training was 0 hours 13 minutes and 53 seconds.
         Total training took 31 epochs.
         The best model was saved in epoch 20 after 4 hours, 45 minutes and 16 seconds.}
         \label{fig:rn1_th_snr1}
     \end{subfigure}
     \hfill
     \begin{subfigure}[b]{0.48\textwidth}
         \centering
         \includegraphics[width=\textwidth]{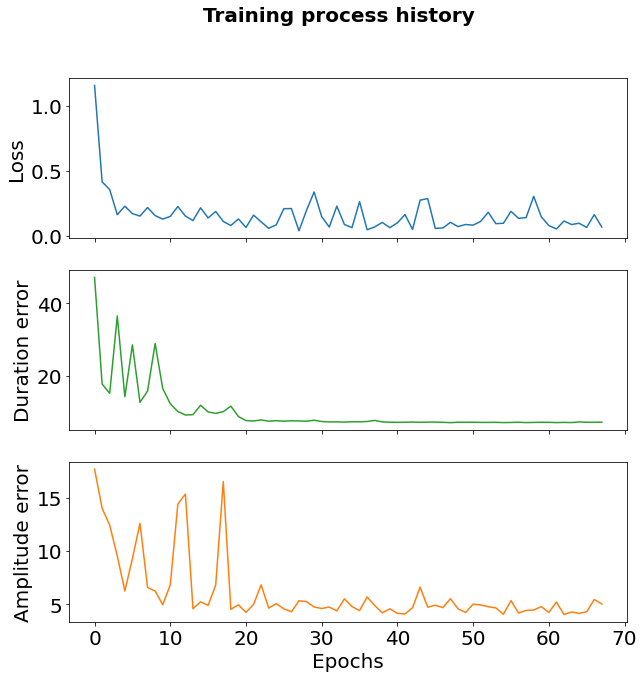}
         \caption{\gls{resnet} 2.
         The total training time was 16 hours, 35 minutes and 50 seconds,
         while the average time during one epoch of training was 0 hours 14 minutes and 38 seconds.
         Total training took 67 epochs.
         The best model was saved in epoch 54 after 13 hours, 27 minutes and 39 seconds.}
         \label{fig:rn2_th_snr1}
     \end{subfigure}
        \caption{Training History trained on the \gls{snr}=1 dataset.}
        \label{fig:bn_th_snr1}
\end{figure}

\begin{figure}[H]
     \centering
     \begin{subfigure}[b]{0.48\textwidth}
         \centering
         \includegraphics[width=\textwidth]{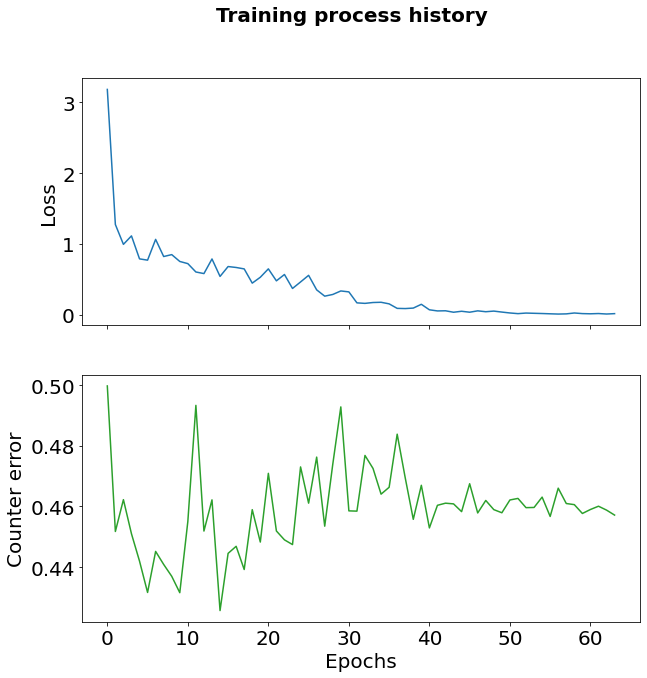}
         \caption{\gls{resnet} 1.
         The total training time was 16 hours, 44 minutes and 52 seconds,
         while the average time during one epoch of training was 0 hours 15 minutes and 42 seconds.
         Total training took 63 epochs.
         The best model was saved in epoch 14 after 3 hours, 54 minutes and 57 seconds.}
         \label{fig:rn1_th_snr0_5}
     \end{subfigure}
     \hfill
     \begin{subfigure}[b]{0.48\textwidth}
         \centering
         \includegraphics[width=\textwidth]{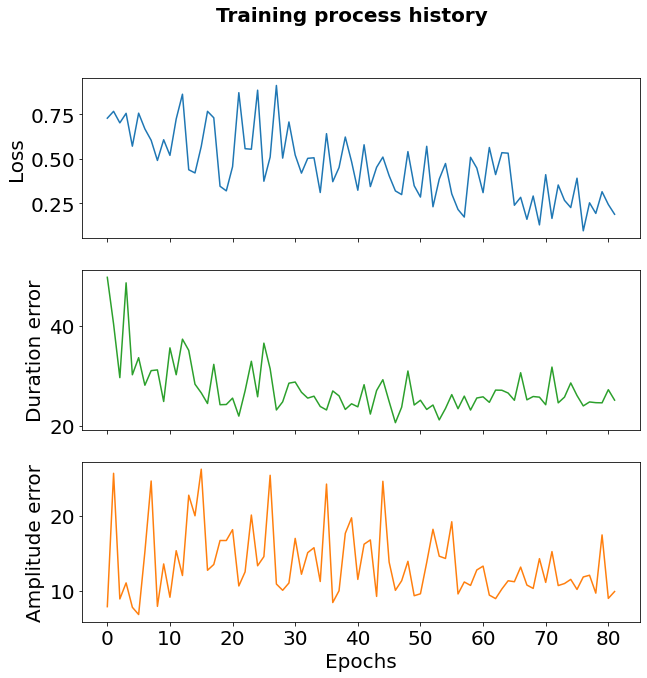}
         \caption{\gls{resnet} 2.
         The total training time was 22 hours, 35 minutes and 16 seconds,
         while the average time during one epoch of training was 0 hours 16 minutes and 31 seconds.
         Total training took 81 epochs.
         The best model was saved in epoch 46 after 12 hours, 55 minutes and 28 seconds.}
         \label{fig:rn2_th_snr0_5}
     \end{subfigure}
        \caption{Training History trained on the \gls{snr}=0.5 dataset.}
        \label{fig:bn_th_snr0_5}
\end{figure}

\begin{figure}[H]
     \centering
     \begin{subfigure}[b]{0.48\textwidth}
         \centering
         \includegraphics[width=\textwidth]{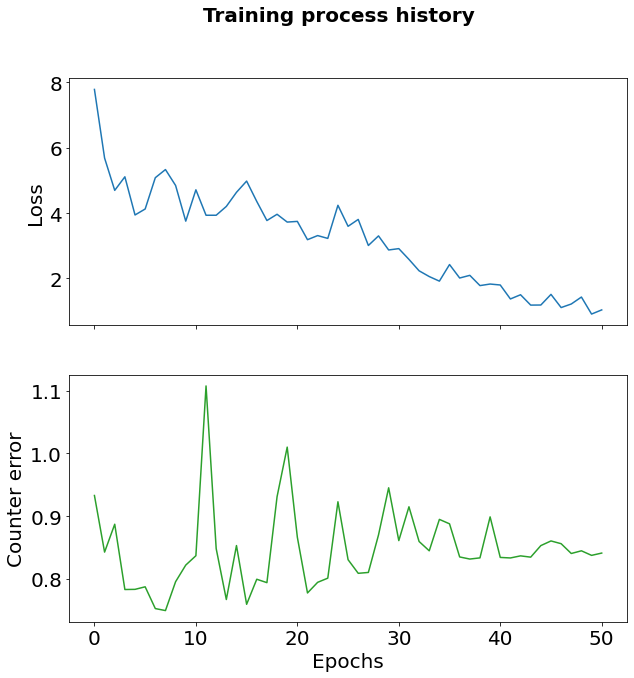}
         \caption{\gls{resnet} 1.
         The total training time was 13 hours, 21 minutes and 16 seconds,
         while the average time during one epoch of training was 0 hours 15 minutes and 42 seconds.
         Total training took 50 epochs.
         The best model was saved in epoch 8 after 2 hours, 9 minutes and 19 seconds.}
         \label{fig:rn1_th_snr0_25}
     \end{subfigure}
     \hfill
     \begin{subfigure}[b]{0.48\textwidth}
         \centering
         \includegraphics[width=\textwidth]{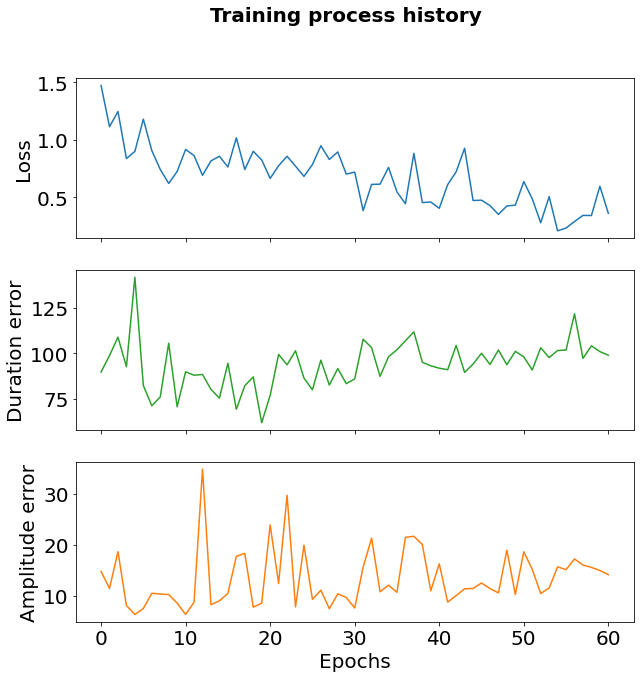}
         \caption{\gls{resnet} 2.
         The total training time was 15 hours, 53 minutes and 1 seconds,
         while the average time during one epoch of training was 0 hours 15 minutes and 37 seconds.
         Total training took 60 epochs.
         The best model was saved in epoch 19 after 5 hours, 21 minutes and 12 seconds.}
         \label{fig:rn2_th_snr0_25}
     \end{subfigure}
        \caption{Training History trained on the \gls{snr}=0.25 dataset.}
        \label{fig:bn_th_snr0_25}
\end{figure}

In \textbf{Figs. \ref{fig:bn_th_snr4}, \ref{fig:bn_th_snr2}, \ref{fig:bn_th_snr1}, \ref{fig:bn_th_snr0_5}} and \textbf{\ref{fig:bn_th_snr0_25}}, we can see the training and validation history for each \gls{bn} instance trained for this work.
Among all the instances, we have each instance trained for a different level of noise (\gls{snr}=4, 2, 1, 0.5 and 0.25).


\newpage

\newpage

\section{Comparison between the results from our neural network and the traditional algorithm}

\begin{figure}[H]
  \centering
  \includegraphics[width=16cm]{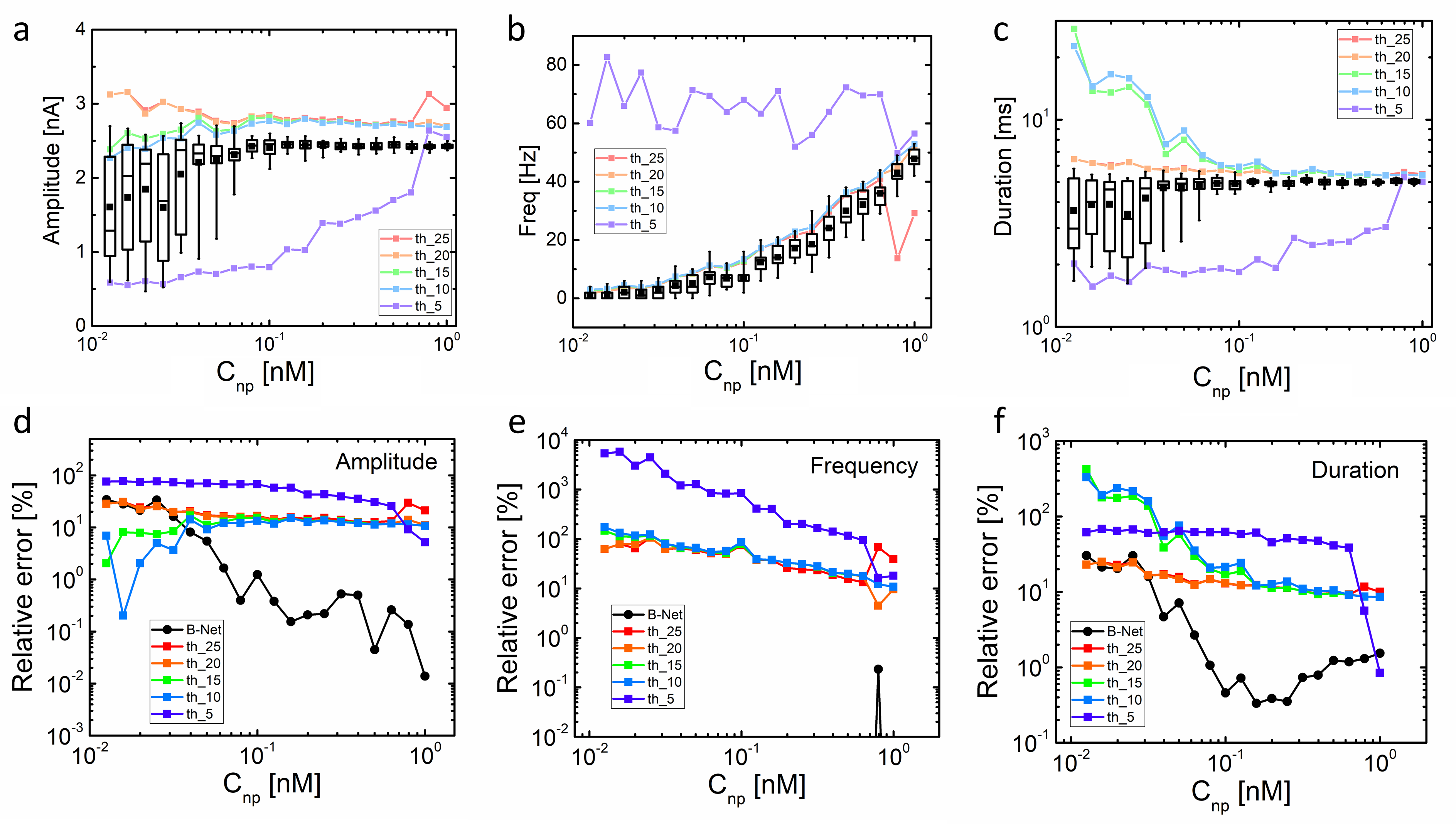}
  \caption{Comparison of the results from our neural network \gls{bn} and those from the traditional algorithm on the dataset of varying \gls{dnp}. (a-c) Box charts of the spike amplitude, frequency, and duration for translocating nanospheres for different \gls{dnp},  respectively, extracted by the \gls{bn}. The average values of the spike amplitude, frequency and duration extracted by the traditional algorithm with different thresholds are shown by the dot-on lines in the corresponding figures for comparison. The relative errors of the spike amplitude, frequency and duration from the \gls{bn} and the traditional algorithm are compared in d-f, respectively.}
  \label{fig:comparizon_X1}
\end{figure}

\begin{figure}[H]
  \centering
  \includegraphics[width=16cm]{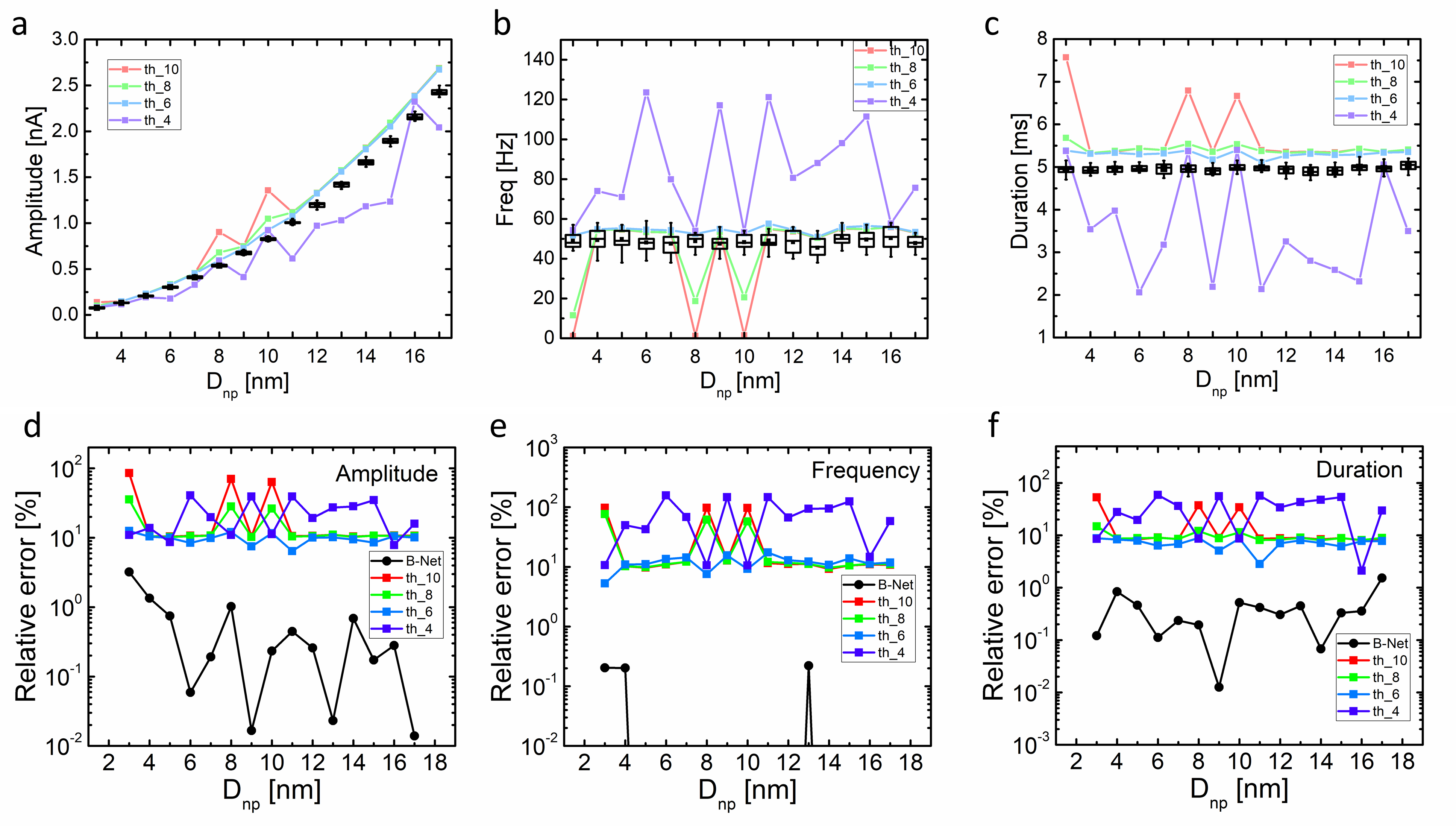}
  \caption{Comparison of the results from our neural network \gls{bn} and those from the traditional algorithm on the dataset of varying \gls{cnp}. (a-c) Box charts of the spike amplitude, frequency and duration for translocating nanospheres for different \gls{cnp},  respectively, extracted by the \gls{bn}. The average values of the spike amplitude, frequency, and duration extracted by the traditional algorithm with different thresholds are shown by the dot-on lines in the corresponding figures for comparison. The relative errors of the spike amplitude, frequency, and duration from the \gls{bn} and the traditional algorithm are compared in d-f, respectively.}
  \label{fig:comparizon_X2}
\end{figure}

\begin{figure}[H]
  \centering
  \includegraphics[width=16cm]{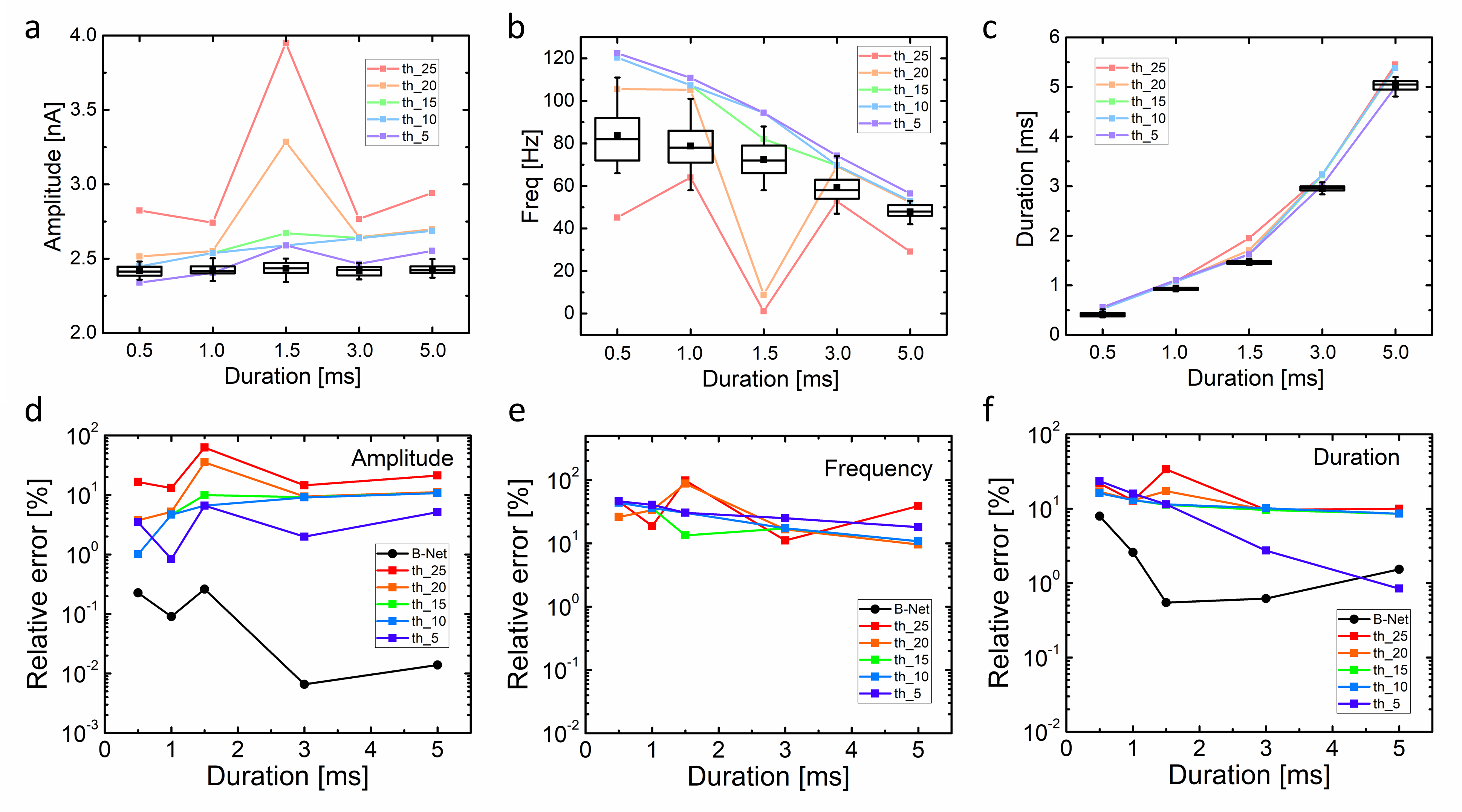}
  \caption{Comparison of the results from our neural network \gls{bn} and those from the traditional algorithm on the dataset of varying duration. (a-c) Box charts of the spike amplitude, frequency and duration for translocating nanospheres for different duration values, respectively, extracted by the \gls{bn}. The average values of the spike amplitude, frequency and duration extracted by the traditional algorithm with different thresholds are shown by the dot-on lines in the corresponding figures for comparison. The relative errors of the spike amplitude, frequency and duration from the \gls{bn} and the traditional algorithm are compared in d-f, respectively.}
  \label{fig:comparizon_X3}
\end{figure}

Enlarging the diameter of translocating nanospheres, increases the amplitude (\textbf{Fig. \ref{fig:comparizon_X1}a}), while increasing the concentration of the nanospheres tend to raise the translocation frequency (\textbf{Fig. \ref{fig:comparizon_X2}b}). The extracted duration agrees well with the set value in the generated signal (\textbf{Fig. \ref{fig:comparizon_X3} (c)}).
However, the average values of these features resulting from the traditional algorithm are dependent on the selection of threshold amplitude, as shown by the deviations among the dot-on lines in the respective figures.
The number $n$ after $th$ (for threshold) in the legend denotes a specific threshold level, measured by the number of multiples of the peak-to-peak value of the background noise.

The prediction errors for the different features for both algorithms are compared in \textbf{d-f} of \textbf{Figs. \ref{fig:comparizon_X1}, \ref{fig:comparizon_X2}} and \textbf{\ref{fig:comparizon_X3}}.
The errors of the \gls{bn} for frequency estimation are almost zero in comparison to the traditional algorithm.
This can be seen in \textbf{e} of \textbf{Figs. \ref{fig:comparizon_X1}, \ref{fig:comparizon_X2}} and \textbf{\ref{fig:comparizon_X3}}, where errors are four orders of magnitude below what the traditional algorithm procedures for several \gls{dnp}, \gls{cnp} and duration values.
The error of the \gls{bn} is up to three orders of magnitude lower in \textbf{d} of \textbf{Figs. \ref{fig:comparizon_X1}, \ref{fig:comparizon_X2}} and \textbf{\ref{fig:comparizon_X3}} and up to two orders of magnitude below in \textbf{f} of \textbf{Figs. \ref{fig:comparizon_X1}, \ref{fig:comparizon_X2}} and \textbf{\ref{fig:comparizon_X3}}.
In addition, the relative errors of the feature extraction for the traditional algorithm are also highly dependent on the selection of amplitude threshold, indicating the subjectivity of the algorithm as well.

\subsection{Artificially generated dataset, varying \gls{dnp}}

\newpage

\subsection{Artificially generated dataset, varying \gls{cnp}}

\newpage

\subsection{Artificially generated dataset, varying duration}

\newpage

\section{Translocation features of $\lambda$-DNA and streptavidin extracted by the B-Net}

\begin{figure}[H]
  \centering
  \includegraphics[width=16cm]{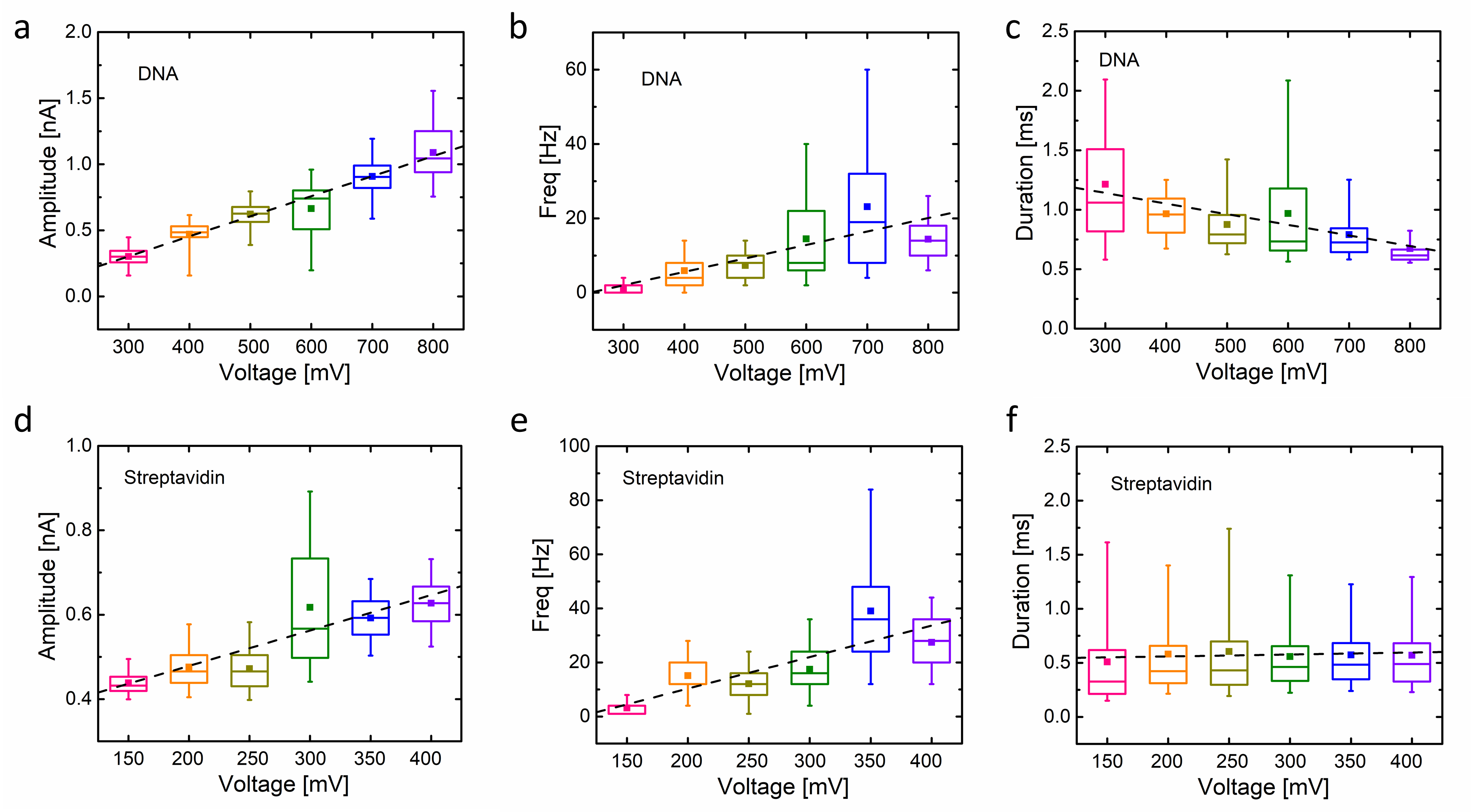}
  \caption{Signal processing of experimental data involving DNA and protein. Variation of spike amplitude, frequency and duration of $\lambda$-DNA (a-c) and \emph{streptavidin} (d-f) translocation with bias voltage. All data are plotted in linear scale and linear regression is applied to each group of data, shown as the black dotted lines.}
  \label{fig:experiment_data}
\end{figure}

\end{appendices}

\bibliographystyle{unsrt}  
\bibliography{References}  